\documentclass[prd, twocolumn, superscriptaddress, showpacs, amsmath, floatfix, byrevtex, preprintnumbers]{revtex4}
\usepackage{graphicx}
\usepackage{amsbsy}
\usepackage{dsfont}

\providecommand{\realline}{\mathds{R}}
\providecommand{\complexnumb}{\mathds{C}}
\providecommand{\prob}{\mathrm{P}}
\providecommand{\normaldistn}{\mathrm{N}}
\providecommand{\imag}{\mathrm{i}}
\providecommand{\expect}{\mathrm{E}}
\providecommand{\var}{\mathrm{Var}}
\providecommand{\invchisq}{\mbox{Inv-}\chi^2}
\providecommand{\re}{\mathrm{Re}}
\providecommand{\im}{\mathrm{Im}}
\providecommand{\solarmass}{\mathrm{M}_{\scriptscriptstyle\odot}}
\providecommand{\transpose}{\prime}
\providecommand{\pz}{\phantom{0}}

\providecommand{\bx}{\boldsymbol{x}}
\providecommand{\bbeta}{\boldsymbol{\beta}}
\providecommand{\bepsilon}{\boldsymbol{\epsilon}}
\providecommand{\bvarepsilon}{\boldsymbol{\varepsilon}}
\providecommand{\bs}{\boldsymbol{s}}
\providecommand{\bm}{\boldsymbol{m}}
\providecommand{\bu}{\boldsymbol{u}}
\providecommand{\bv}{\boldsymbol{v}}

\providecommand{\by}{\boldsymbol{y}}
\providecommand{\bX}{\boldsymbol{X}}
\providecommand{\bH}{\boldsymbol{\mathrm{H}}}
\providecommand{\bU}{\boldsymbol{\mathrm{U}}}
\providecommand{\bV}{\boldsymbol{\mathrm{V}}}
\providecommand{\bS}{\boldsymbol{\mathrm{S}}}
\providecommand{\pdf}{\mathrm{p}}
\providecommand{\bmu}{\boldsymbol{\mu}}
\providecommand{\bSigma}{\boldsymbol{\Sigma}}
\providecommand{\exA}{\textsl{A}}
\providecommand{\exB}{\textsl{B}}
\providecommand{\exC}{\textsl{C}}

\newenvironment{equationarray}
{\arraycolsep 0.14 em
\begin{eqnarray}}
{\end{eqnarray}}

\DeclareMathOperator{\diag}{diag}
\DeclareMathOperator{\rank}{rank}
\DeclareMathOperator{\tr}{tr}

\begin{document}
  \title{Bayesian reconstruction of gravitational wave burst signals \\
    from simulations of rotating stellar core collapse and bounce}
  \preprint{\texttt{LIGO-P0900089}}
  \preprint{\texttt{AEI-2009-089}}
  \date{\today}
  \author{Christian R\"{o}ver}
  \affiliation{Max-Planck-Institut f\"{u}r Gravitationsphysik
    (Albert-Einstein-Institut), 30167~Hannover, Germany}
  \author{Marie-Anne Bizouard}
  \affiliation{Laboratoire de l'Acc\'{e}l\'{e}rateur Lin\'{e}aire,
    Universit\'{e} Paris Sud, 91898~Orsay, France}
  \author{Nelson Christensen}
  \affiliation{Physics and Astronomy, Carleton College, Northfield,
    Minnesota~55067, USA}
  \author{Harald Dimmelmeier}
  \affiliation{Department of Physics, Aristotle University of
    Thessaloniki, 54124~Thessaloniki, Greece}
  \author{Ik Siong Heng}
  \affiliation{Department of Physics and Astronomy, University of
    Glasgow, Glasgow~G12~8QQ, UK}
  \author{Renate Meyer}
  \affiliation{Department of Statistics, The University of Auckland,
    Auckland~1142, New Zealand}


\begin{abstract}
  Presented in this paper is a technique that we propose for
  extracting the physical parameters of a rotating stellar core
  collapse from the observation of the associated gravitational wave
  signal from the collapse and core bounce.
  Data from interferometric gravitational wave detectors can be used
  to provide information on the mass of the progenitor model,
  precollapse rotation and the nuclear equation of state.
  We use waveform libraries provided by the latest numerical
  simulations of rotating stellar core collapse models in general
  relativity, and from them create an orthogonal set of eigenvectors
  using principal component analysis.
  Bayesian inference techniques are then used to reconstruct the
  associated gravitational wave signal that is assumed to be detected
  by an interferometric detector. Posterior probability distribution
  functions are derived for the amplitudes of the principal component
  analysis eigenvectors, and the pulse arrival time.
  We show how the reconstructed signal and the principal component
  analysis eigenvector amplitude estimates may provide information on
  the physical parameters associated with the core collapse event.
\end{abstract}

\pacs{02.70.Uu, 04.80.Nn, 05.45.Tp, 97.60.Bw}

\maketitle


\section{Introduction}
\label{section:introduction}

The detection of gravitational radiation will likely come in the near
future. LIGO~\cite{AbramoviciEtAl1992, Barish1997, AbbottEtAl2004a} is
at its initial target sensitivity, and the detection of an event from
an astrophysical source could come at any time. Around the globe a
world-wide network of detectors is emerging; VIRGO in
Italy~\cite{CaronEtAl1996, Brillet1997}, GEO in
Germany~\cite{HoughEtAl1997} and TAMA in Japan~\cite{TsubonoEtAl1997}
are operating alongside the LIGO detectors in the U.S.\ in the quest
for gravitational wave detection. In a few years \textsl{advanced
LIGO} and \textsl{advanced Virgo} will come on-line, with a
sensitivity increase of 10 over initial LIGO~\cite{SmithEtAl2009,
  advLigoURL, AcerneseEtAl2008, advVirgoURL}, increasing the
prospective detection rate significantly. It will be a great day for
physics when gravitational waves are finally directly detected, but it
will also be the birth of a new way of observing the universe and
conducting astronomy and astrophysics.

The observations of expected sources should be dramatic: stellar
collapse, pulsars, the inspiral of binary neutron stars followed by
black hole formation, or even the stochastic background from the Big
Bang. Gravitational wave astronomy will soon enter a regime where
parameter estimation work will provide the means to make important
astrophysical statements. Gravitational wave burst signals from
rotating stellar core collapse and bounce, resulting in the formation
of a proto-neutron star, are one of the more promising and potentially
extremely valuable sources for ground based interferometric detectors.
Gravitational wave burst events are typically characterized by their
short duration (from a few milliseconds to about one second) and the
absence of accurate theoretical predictions for their waveforms (as
opposed to e.g.\ waveform templates for binary inspiral). If these
events happen sufficiently close by, the ground based gravitational
wave detectors will be able to observe them. Detection searches for
gravitational wave bursts typically use methods that can identify
unmodeled but short duration events~\cite{BeauvilleEtAl2008}.
LIGO~\cite{AbbottEtAl2007, AbbottEtAl2008a, AbbottEtAl2005c},
VIRGO~\cite{AcerneseEtAl2009}, GEO~\cite{AbbottEtAl2008a}, and
TAMA~\cite{AbbottEtAl2005c, AndoEtAl2005b} have all recently conducted
searches for gravitational wave bursts.

The prediction of the exact burst signal to be expected from a
rotating stellar core collapse event depends on the complex interplay
of general relativity, nuclear and particle physics. 
Furthermore, it is anticipated that the signal is produced by various 
emission mechanisms, first from the coherent motion of the collapsing 
and rebounding  core during the proto-neutron star formation and then 
the ringing down of the nascent hot proto-neutron star (all in cases 
when the core rotates). This is then possibly followed by emission 
due to prompt convective motion behind the hydrodynamic shock in the 
central part of the star due to non-axisymmetric rotational 
instabilities for rapid rotation, as well as due to pulsations of the 
proto-neutron star, for instance, triggered by fall-back of matter 
onto it (see e.g.~\cite{Ott2009} for a comprehensive review).
Only recently, the
full complexity of the prospective emission mechanisms for
gravitational waves in a stellar core collapse has become appreciated.
Previously, research mainly concentrated on the signal contribution
from the collapse, core bounce and ring-down phase, and therefore this
is the gravitational wave signal that is now by far most
comprehensively and accurately investigated using numerical studies.
Recently, a series of calculations of core collapse models with
unprecedented physical realism has been performed~\cite{OttEtAl2007,
DimmelmeierEtAl2007, DimmelmeierEtAl2008a}. 
These studies provide strong evidence that the predictions of 
gravitational wave signals from the collapse and bounce phase are now 
robust. The gravitational wave results from these investigations have 
been made publicly available in the form of electronic waveform 
catalogs.
If detected,
such a gravitational wave signal can then ideally provide information
on the mass of the progenitor model, the precollapse rotation, the
density of the collapsed core, and the nuclear equation of state.

However, even for the collapse and core bounce signal (not to mention
the possible signal contributions from other emission mechanisms in a
stellar core collapse event) one cannot conduct a template based
search as is done when looking for coalescing compact binary signals,
since it would be computationally impossible to completely cover the
signal parameter space, and therefore various other methods have been
developed to efficiently perform the search for gravitational wave
burst signals in the detectors. Bayesian inferential methods provide a
means to use data from interferometric gravitational wave detectors in
order to extract information on the physical parameters associated
with an event~\cite{Finn1997}. Markov chain Monte Carlo (MCMC) methods
are a powerful computation technique for parameter estimation within
this framework; they are especially useful in applications where the
number of parameters is large~\cite{MCMCinPractice}. Good descriptions
of the positive aspects of a Bayesian analysis of scientific and
astrophysical data are provided in~\cite{Finn1997, Loredo1992,
  Gregory2001}. MCMC routines have been developed that will produce
parameter estimates for gravitational wave signals from binary
inspiral~\cite{ChristensenMeyer2001, ChristensenMeyerLibson2004,
  RoeverMeyerChristensen2006a, RoeverMeyerChristensen2007a,
  RoeverEtAl2007b, VanDerSluysEtAl2008b} and
pulsars~\cite{ChristensenEtAl2004, UmstaetterEtAl2004,
  VeitchEtAl2005}. Accurate parameter estimation of observed
gravitational radiation events is the necessary pathway to
understanding important problems in astrophysics and cosmology. In
this paper we present methods which are the starting point in the
process of using MCMC methods to extract and estimate parameters
associated with rotating stellar core collapse events from their
gravitational wave signals emitted during the collapse, core bounce
and ring-down phase.

Unfortunately, due to the complexity of the event (in particular its
intrinsic multi-dimensional nature), the computational
time required to derive these signals in numerical simulations of
rotating stellar core collapse is significant, and thus the waveform
generation cannot be performed instantly  while fitting a signal
template to the data, and additional techniques to simplify the
analysis are required. Instead of using waveforms corresponding to
arbitrarily picked locations within parameter space, we reduce the
complexity of the problem by simplifying the \textsl{waveform space}
to the span of a small number of basis vectors. The basis vectors are
derived from a representative catalog of numerically computed signal
waveforms through the use of \textsl{principal component analysis}
(PCA)~\cite{Heng2009}. The waveforms used in our present analysis are
from the most recent, advanced, and comprehensive general relativistic
study of rotating stellar core collapse by Dimmelmeier et
al.~\cite{DimmelmeierEtAl2008a} and depend on the mass of the
progenitor model, the precollapse rotation, and the nuclear equation
of state.

In this study we make use of a catalog that is smaller than
eventually required in an extensive and accurate analysis of signals
in real data, but with its non-trivial number of elements spanning a
large portion of the interesting parameter space we expect it
nevertheless to be sufficiently complex to probe and to demonstrate
the method qualitatively. The MCMC algorithm essentially fits a
superposition of derived basis vectors to the data, providing
parameter estimates as well as associated uncertainties. We also demonstrate how
PCA and MCMC then allow the reconstruction of the time series of a
gravitational wave burst signal, including confidence bands. We
present initial results showing that these PCA basis vectors are
actually physically meaningful, and that we can extract astrophysical
parameters from the measured core collapse burst signal. The ultimate
goal is to be able to extract information on the nuclear equation of
state from the gravitational wave signal as it is observed by a
network of interferometers, as well as other relevant physical
parameters associated with the event, as far as this is possible
from the collapse, bounce and ringdown signal alone.

There have been other approaches to signal reconstruction and
parameter estimation with burst signals from stellar core
collapse. The discipline essentially starts with the work of
G\"{u}rsel and Tinto~\cite{GuerselTinto1989}, who presented a method
for reconstructing the time series for gravitational wave bursts.
Rakhmanov~\cite{Rakhmanov2006} developed a general scheme of Tikhonov
regularization to be applied on a network of detectors in order to
extract the signal. MCMC methods have also been applied to parameter
estimation for burst events from cosmic string
cusps~\cite{KeyCornish2009}. Searle et
al.~\cite{SearleSuttonTinto2009} have recently worked out the Bayesian
framework to the coherent detection and analysis of burst signals,
including a wide range of special cases like white noise burst
signals,  but also signals constituting linear combinations of some
set of basis vectors, like the ones used in this present study.
Summerscales et al.~\cite{SummerscalesEtAl2008} proposed a maximum
entropy technique in order to infer the time-dependent gravitational
wave burst signals detected by a network of interferometric detectors;
they then calculated the correlation between their reconstructed
waveform and the entries in a catalog of rotating core collapse
waveforms~\cite{OttEtAl2004} in order to infer the physical parameters
from the event. Our work is different in that we use waveforms from
physical models (the table of rotating core collapse signals) as our
means of reconstructing the signal. PCA basis vectors are initially
created from the table of waveforms, and the MCMC estimate of the
amplitudes for the PCA basis vectors provides the reconstructed
waveform. A linearly polarized gravitational wave signal can be
reconstructed from the data of a single interferometer. The estimate
of the physical parameters comes from an association of the
reconstructed waveform with the elements of the catalog. The results
in this paper are derived only using data from a single detector
(signals embedded in simulated noise matching that of LIGO at its
target sensitivity).

The organization of the paper is as follows. In
Section~\ref{section:pipelines} we describe the methods by which LIGO,
VIRGO, GEO, and TAMA are searching for burst events; the parameter
estimation technique we describe in this paper would be applied to
event candidates from such a  burst search. The method by which
the catalog of stellar core collapse burst signals is established is
described in Section~\ref{section:models}. A summary of the PCA
method used in our study is presented in Section~\ref{section:pca},
while our analysis strategy is described in
Section~\ref{section:analysis}. We illustrate how the analysis method
works when applied to simulated example data in
Section~\ref{section:implementation}, and in
Section~\ref{section:correspondence} we demonstrate that the PC basis
vectors and the inferred corresponding coefficients actually carry
physically meaningful information. In
Section~\ref{section:conclusions} we provide a summary, and discuss
the direction of future work.


\section{Burst search pipelines}
\label{section:pipelines}

All collaborations associated with ground based interferometric
gravitational wave detectors have developed burst search pipelines in
order to look for events~\cite{AbbottEtAl2007, AbbottEtAl2008a,
  AbbottEtAl2005c, AcerneseEtAl2009, AndoEtAl2005b}. Although some
gravitational wave burst sources are well modeled, the burst pipelines
search for very general types of waveforms. The only assumption which is usually
made concerns the duration of the signal (less than a few hundred of
ms) and the frequency bandwidth of the search, which can be as large as
the detector bandwidth (from a few dozen of Hz up to the Nyquist
frequency). Burst pipelines typically reduce the frequency bandwidth to
a few kHz to focus on specific types of sources. 

Most of the gravitational wave burst pipelines look for an excess of
energy in a time--frequency map using different multi-resolution
time--frequency transforms. Anderson et al.~\cite{AndersonEtAl2001}
initially considered the energy given by the Fourier transform in a
frequency band. More recent methods~\cite{KlimenkoMitselmakher2004} 
make use of a wavelet decomposition of the data stream.
In contrast, in~\cite{ChatterjiEtAl2004, Chatterji2005, ClapsonEtAl2008} 
a gravitational wave burst signal represented as a sine-Gaussian is implemeted.

Usually the gravitational wave burst searches are performed using data
from a network of detectors. Demanding that a gravitational wave burst
event is seen simultaneously in several detectors allows one to reduce
the false alarm rate, which is rather high in a burst search due to
the short duration of the signal. This is also the only way to
disentangle a real gravitational wave signal from transient noise
events. There exist two kinds of network analysis: coincident or
coherent filtering. In a coincidence analysis~\cite{ChatterjiEtAl2004,
 ClapsonEtAl2008}, each interferometer output is analyzed, providing
a list of triggers. A coincidence in time and frequency is then
required. A coherent analysis~\cite{GuerselTinto1989, Sylvestre2003,
 ArnaudEtAl2003, WenSchutz2005, MohantyRakhmanovKlimenkoMitselmakher2006, 
KlimenkoYakushinMercerMitselmakher2008, Sutton2009} uses all 
interferometers' information by combining the input data streams or 
the filtered data streams into one single output which takes into account 
the detectors' antenna response function assuming the source is at a 
given position in the sky. A coherent analysis more efficienctly 
suppresses non-stationarities that are expected to be incoherent in the 
different detectors.

When using data from a network of detectors, burst pipelines can
reconstruct the position in the sky of the
source~\cite{GuerselTinto1989, CavalierEtAl2006, MarkowitzEtAl2008,
 Fairhurst2009}. The accuracy of the source sky position depends
on the time resolution of the pipeline. It is usually estimated for a
set of different waveforms. Depending on the complexity of the signal
waveform the time resolution can vary from a fraction of ms up to
several ms~\cite{BeauvilleEtAl2008}. The frequency content (central frequency
and bandwidth) of the event is also estimated by the burst
pipelines. The frequency information can give some hints concerning
the possible astrophysical source. For instance, a central
frequency of a few hundred Hz would point towards the event of a
core collapse in a massive star resulting in the formation of a
proto-neutron star~\cite{DimmelmeierEtAl2008a} (in particular if the
detection is accompanied by a neutrino trigger following shortly
after) while a higher frequency content at about $ 10 \mathrm{\ kHz} $
could suggest that the event could be due to a neutron star
collapsing to a black hole~\cite{BaiottiEtAl2005}. Besides, coherent
pipelines can also extract from the data an estimation of the
waveform without assuming any model for the triggers with sufficient
amplitude~\cite{GuerselTinto1989, FlanaganHughes1998b}.

The work presented in this paper intends to extract even more
information from the most significant events found by a gravitational
wave burst pipeline assuming an astrophysical source model. The
estimation of the parameters could also be used to reject a possible
gravitational wave event candidate;
the idea of distinguishing a real stellar core collapse signal from
an instrumental \textsl{glitch} is briefly discussed in
sections~\ref{section:correspondence} and \ref{section:conclusions}
below.
The starting point for our technique would be the list of candidate 
triggers produced by a burst all-sky search pipeline, or times provided 
by electromagnetic and/or neutrino observations (external triggers). 
In both cases, the MCMC will search for the event over some relatively 
small time span. The trigger times provide a relatively small number 
of data periods to be examined by our method. The MCMC would attempt 
to produce a reconstruction of the signal based on a waveform catalog 
for stellar core collapse models.


\section{Gravitational wave signals from rotating core collapse and
  bounce}
\label{section:models}

In stellar core collapse, the (possibly rotating) unstable iron core
of a massive star at the end of its life contracts to supernuclear
density on a dynamical time scale of the order of $ 100 \mathrm{\ ms} $.
As the core material stiffens due to repulsive nuclear forces, the
collapse is halted abruptly, and the inner part of the core undergoes
a rebound (the core bounce). After a period of ring-down oscillations,
the hot proto-neutron star settles down, while the remainder of the
star is possibly (mainly depending on the mass of the progenitor)
blown off in a supernova explosion.

If the precollapse stellar core is rotating, the dynamics of the
evolution are reflected in the gravitational wave signal waveform with
a slow rise during the core collapse, a large negative peak around
bounce, and damped oscillations in the ring-down phase. If
rotational effects become important for rapidly rotating models, the
core collapse can even be stopped by centrifugal forces alone at
subnuclear density. However, in contrast to previous, less
sophisticated studies of the stellar core collapse scenario (see
e.g.~\cite{DimmelmeierEtAl2002b} and references therein), Dimmelmeier
et al.~\cite{DimmelmeierEtAl2007} have shown that the signal
waveform remains qualitatively unaltered and is thus generic for a
wide range of initial rotation strength.

The gravitational radiation waveforms analyzed in this article are the
burst signals from the most recent, advanced, and comprehensive
general relativistic study of rotating stellar core collapse
models~\cite{DimmelmeierEtAl2008a}. These simulations were performed
with a computational code that utilizes accurate Riemann solvers to
evolve the general relativistic hydrodynamic equations and a nonlinear
elliptic solver based on spectral methods~\cite{DimmelmeierEtAl2005}
for the fully coupled metric equations in the conformal flatness
approximation of general relativity~\cite{Isenberg2008}. The
precollapse iron core models were taken from recent stellar
evolutionary calculations by Heger et al.~\cite{HegerLangerWoosley2000,
  HegerWoosleySpruit2005}, either with an intrinsic or an artificially
added rotation profile. These initial models were then evolved with a
nonzero-temperature nuclear equation of state (EoS), either the one by
Shen et al.~\cite{ShenEtAl1998a, ShenEtAl1998b} (Shen EoS) or the one
by Lattimer and Swesty~\cite{LattimerEtAl1985, LattimerSwesty1991} (LS
EoS; with a bulk incompressibility $ K = 180 \mathrm{\ MeV} $), both
in the implementation of Marek et al.~\cite{MarekEtAl2005},
including contributions from baryons, electrons, positrons, and
photons. Deleptonization by electron capture on nuclei and free
protons during the collapse phase is realized as proposed and tested
by Liebend\"orfer~\cite{Liebendoerfer2005}.

Of the 136 models investigated in the study by Dimmelmeier et
al.~\cite{DimmelmeierEtAl2008a}, 128 models have an analytic initial
rotation profile. Their collapse behavior is determined by the
following parameters:
\begin{itemize}
\item The strength of rotation is specified by the \textsl{precollapse
    central angular velocity}, which varies from
  $ \Omega_\mathrm{c,i} = 0.45 $ to $ 13.31 \mathrm{\ rad\ s}^{-1} $
  (with the individual range depending on the differentiality of
  rotation). The influence of rotation strength on the collapse
  dynamics and on the burst signal via rotational flattening is
  very pronounced.

  For a wide range of slow to intermediate initial rotation
  strengths, the peak $ |h|_\mathrm{max} $ of the gravitational wave
  amplitude is almost proportional to the ratio
  $ T_\mathrm{b} / |W|_\mathrm{b} $ of rotational energy to
  gravitational energy at bounce (which in turn increases
  approximately linear with $ \Omega_\mathrm{c,i} $ in this regime if
  all other model parameters are kept constant).

  For very rapid rotation, however, $ |h|_\mathrm{max} $ reaches a
  maximum and declines again, when the centrifugal barrier slows down
  the contraction considerably, preventing the core from collapsing to
  high supernuclear density. In such a case the frequency of the burst
  signal, which is practically constant for slow or moderate rotation,
  decreases significantly. Note that the trend to lower frequencies for
  very slowly rotating models shown in~\cite{DimmelmeierEtAl2007,
    DimmelmeierEtAl2008a}, which spoils the constancy of the signal
  frequency for such models, is due to a (possibly artificial)
  low-frequency contribution originating from post-bounce convection
  in the proto-neutron star, which superimposes the signal from the
  core bounce and proto-neutron star ring-down.
\item The \textsl{differentiality of the precollapse rotation profile}
  is set by a length scale with values $ A = 50\,000 $, $ 1\,000 $, or
  $ 500 \mathrm{\ km} $, ranging from almost uniform to strongly
  differential rotation. For a fixed initial strength of rotation
  (as set by $ \Omega_\mathrm{c,i} $), a change in $ A $ alone has no strong
  effect on the gravitational wave signal, neither on the amplitude
  nor on the frequency. However, rapid rotation can only be achieved
  for comparably small values of $ A $.
\item The set of precollapse cores encompasses models with a
  \textsl{progenitor mass} of $ M_\mathrm{prog} = 11.2 $, $ 15.0 $,
  $ 20.0 $, or $ 40.0 \, \solarmass $ (masses at zero-age main
  sequence). The choice of the progenitor mass has a direct impact on
  the mass of the inner core during the collapse and at bounce (and
  thus on the mass of the proto-neutron star). Already without
  rotation, the different progenitors produce an inner core at
  bounce with a mass that depends non-monotonously on the mass of the
  progenitor (with the inner core mass increasing in the order
  $ M_\mathrm{prog} = 11.2 $, $ 20.0 $, $ 15.0 $, and
  $ 40.0 \, \solarmass $).

  This variation is considerably amplified by rotation, which itself
  increases the mass on the inner core at bounce (approximately
  linear with $ \Omega_\mathrm{c,i} $ at slow rotation and roughly
  quadratically at rapid rotation; see~\cite{DimmelmeierEtAl2008a}).
  Nevertheless, the reflection of that effect on the peak
  signal amplitude $ |h|_\mathrm{max} $ is practically negligible, as
  the effect of a high inner core mass (which causes a large
  quadrupole moment and thus a strong resulting gravitational wave
  signal) is canceled by strong centrifugal support, resulting in
  slower collapse dynamics and, consequently, a weaker gravitational
  wave signal.

  The main frequency of the gravitational wave burst signal is also
  not influenced significantly 
  by the mass of the progenitor model,
  except that in models with the least massive progenitor
  ($ M_\mathrm{prog} = 11.2 \, \solarmass $) even strong initial rotation
  does not decelerate the relatively small collapsing inner core
  enough to make it bounce at subnuclear densities. Hence the signal
  frequencies of such models remain comparably high.
  We also point out that the approximate nature of the deleptonization
  scheme during the collapse phase up to core bounce employed in these
  models could actually be responsible for overemphasizing the
  variation in the mass of the inner core at bounce with respect to
  the mass of the progenitor.
\item The \textsl{microphysical equation of state during the
    evolution} is chosen to be either the Shen EoS or the LS EoS. In the
  subnuclear density regime, the two EoSs are rather similar, with the
  LS EoS being a bit softer on average. However, at supernuclear
  density, the differences are more pronounced. Here the adiabatic index
  $ \gamma $ of the LS EoS jumps to $ \sim 2.5 $, while in the Shen
  EoS $ \gamma $ reaches values of $ \sim 3.0 $, making the nuclear
  material described by that EoS significantly stiffer. Consequently,
  the models with the Shen EoS consistently exhibit lower central
  densities at bounce and after ring-down.

  Still, with respect to the peak waveform amplitude $ |h|_\mathrm{max} $ of the
  burst signal, this does not translate into an unequivocal trend due
  to a complicated interplay in the proto-neutron star between central
  compactness and density structure at intermediate radii. The peak
  frequency of the waveform spectrum, on the other hand, is almost
  always lower if the Shen EoS is used, which is a direct consequence
  of the lower central densities in the bounce and post-bounce period
  due to the nuclear material stiffness from that EoS if all other parameters are
  identical and only the EoS is varied.
\end{itemize}

\begin{figure}
  \includegraphics[width=\columnwidth]{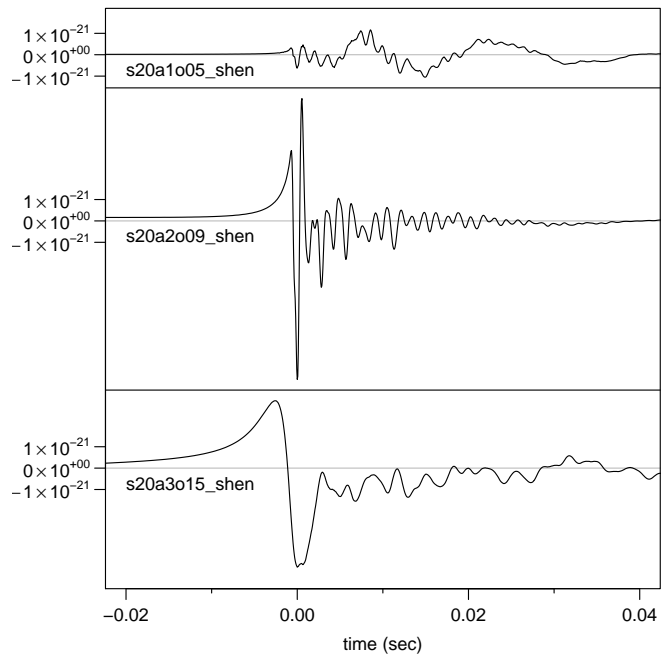}
  \caption{Sample waveforms of the core bounce signal for three models with
           varying initial rotation states while the mass of the progenitor model
           and EoS are fixed. Note the relatively small signal peak at the
           time of core bounce and the significant late-time contribution
           from post-bounce convection for the slowly rotating model
           ``\texttt{s20a1o05\_shen}'', and the overall lower signal frequency 
           for the rotation-dominated and centrifugally bouncing model 
           ``\texttt{s20a3o15\_shen}''. The three signals presented here 
           adequately cover the waveform morphology of our signal catalog.}
  \label{fig:CatalogExamples}
\end{figure}
In Fig.~\ref{fig:CatalogExamples} we present a representative sample of
waveforms which illustrate the impact of a model's
initial rotation state on the gravitational wave signal
from core bounce, from slow and almost uniform initial
rotation (model ``\texttt{s20a1o05\_shen}''),  moderately
fast and modestly differential initial rotation (model
``\texttt{s20a2o09\_shen}''), to rapid and very differential initial
rotation (model ``\texttt{s20a3o15\_shen}'').
While the effect of varying the initial rotation state is clearly
reflected in the waveform for the models shown here, the influence
of the progenitor model's mass and also the EoS for a fixed initial
rotation state is much less apparent, and thus analogous pictorial
waveform comparisons are not presented. 

We note that 8 progenitor models of~\cite{DimmelmeierEtAl2008a} are
from a stellar evolutionary calculation that includes rotation in an
approximate way. Hence, these models already have an initial angular
velocity profile; no artificial rotation according to the simple
analytic relation, like in the other models, is added prior to
evolution. Consequently, as initial rotation is not parametrized,
these models are not of use for this study. However, the collapse
dynamics and associated signal waveforms of these models are well
represented by models with artificially added precollapse rotation in
terms of both signal amplitude and frequency, and therefore it is
justified to not separately consider their behavior here. The
different influences of the various model parameters, which are
summarized here only briefly, are discussed in detail
in~\cite{DimmelmeierEtAl2008a}. The respective signal data can
be downloaded freely from an online waveform template
catalog~\cite{DimmelmeierWaveCatalogue}.

We emphasize that the parameter selection for the models
in~\cite{DimmelmeierEtAl2008a} is fairly complete in that it
accounts for all known relevant parameters which could have an impact
on the burst signal from a core bounce (although, for instance in the
case of neutrino effect, in an approximate way). Furthermore, the
astrophysically meaningful range of the parameters has been reasonably
exhausted (in the case of rotation, the expected strength of rotation
for the bulk of stellar core collapse events is even at the lower end
of the investigated range; see~\cite{WoosleyEtAl2006}). Only the
selection of EoS is limited to two due to a lack of access to
alternative nonzero-temperature nuclear equations of state for stellar
core collapse at the time when the study by Dimmelmeier et
al.~\cite{DimmelmeierEtAl2008a} was performed. However, with the Shen
EoS and the LS EoS the two extremes of a rather stiff nuclear material
and a somewhat soft one are probably well covered.

The range of frequencies and amplitudes of the signals is quite
broad for the model sample of~\cite{DimmelmeierEtAl2008a}, with the
\textsl{integrated characteristic frequency} $ f_\mathrm{c} $ spanning
from about $ 100 $ to $ 700 \mathrm{\ Hz} $ (as opposed to the peak
frequency $ f_\mathrm{max} $ of the spectrum which lies in a narrow
interval around $ 700 \mathrm{\ Hz} $; see Fig.~16
in~\cite{DimmelmeierEtAl2008a}) and the \textsl{integrated
  dimensionless characteristic} amplitude $ h_\mathrm{c} $ ranging
from $ 6 \times 10^{-22} $ to $ 7 \times 10^{-21} $ for e.g.\ initial
LIGO and a distance to the source of $ 10 \mathrm{\ kpc} $ (for the
definition of these quantities, see the references in
e.g.~\cite{DimmelmeierEtAl2008a}).

Still, the main factor of influence causing this variation is the
rotation of the core (as measured by $ \Omega_\mathrm{c,i} $ and
$ A $, and rather well resolved in the models
of~\cite{DimmelmeierEtAl2008a}), while the mass of the progenitor,
precollapse differentiality of rotation, and microphysical EoS hardly
affect both $ f_\mathrm{c} $ and $ h_\mathrm{c} $ (see Fig.~17
in~\cite{DimmelmeierEtAl2008a}). This partial degeneracy regarding the
model parameters makes it very difficult to infer the unknown
parameters from a detection of the gravitational wave burst
signal emitted by a rotating stellar core collapse. In reverse, this means
that parameters of little impact on the gravitational wave signal may
be resolved only coarsely in a parameter study aiming at providing
signal templates.

For an event with at most moderate rotation, in practice only the
strength of rotation can be extracted from the waveform with confidence,
provided the distance to the source and the orientation with respect
to the rotation axis can also be determined. Only if a multitude of
core collapse events can be detected via gravitational waves can
systematic effects of e.g.\ the nuclear material stiffness 
(as described by the EoS) on the frequency also be analyzed with more
certainty.

While the independent parameters $ \Omega_\mathrm{c,i} $, $ A $,
$ M_\mathrm{prog} $, and EoS uniquely specify each collapse model,
from the actual numerical simulation of the collapse a number of
important quantities can be obtained which characterize the evolution
of each individual model; these are typically presented as results for
rotating stellar core collapse calculations. Among these are the
maximum density $ \rho_\mathrm{max,b} $ at the time of core bounce,
the ratio $ T_\mathrm{b} / |W|_\mathrm{b} $ of rotational energy to
gravitational energy at bounce and the corresponding value
$ T_\mathrm{pb} / |W|_\mathrm{pb} $ late after bounce.

These quantities provide information about the collapse dynamics and,
for instance, permit one to distinguish between a core bounce that is
mostly caused by the nuclear material stiffening at supernuclear
densities to one that is dominated by centrifugal forces, which is
again reflected by the waveform. Thus, in reverse the waveform encodes
not only information about the independent model parameters, but also
about these evolution quantities. Consequently, in the analysis on the
correspondence between PCs and physical parameters presented in
Section~\ref{section:correspondence}, we not only consider the
parameters for the model setup but also the evolution of the
quantities $ \rho_\mathrm{max,b} $, $ T_\mathrm{b} / |W|_\mathrm{b} $,
and $ T_\mathrm{pb} / |W|_\mathrm{pb} $. However, in contrast to these
`robust' parameters reflecting the global collapse dynamics we refrain
from analyzing other quantities (like e.g.\ entropy at a specific
off-center location or the time span of contraction) which are prone
to depend more sensitively on the numerical evolution scheme or grid
setup used for the model simulation.


\section{Singular value decomposition}
\label{section:pca}

The signal waveforms used in the analysis reported here are
originally generated at a higher sampling rate. Each waveform is
subsequently resampled at a rate of $16\,384$~Hz (the LIGO and
GEO data sampling rate). The waveforms are then buffered with zeros so
that they are of the same length. Finally, the waveforms are time
shifted so that the first (negative) peak in each waveform (which
occurs shortly after the time of core bounce) are aligned.

For modeling purposes, the set of signal waveforms will be decomposed
into an orthonormal basis. Following the method prescribed
by~\cite{Heng2009}, we create a matrix $ \bH $ so that each column
corresponds to a signal waveform from the catalog after subtracting
the overall mean of the waveforms. For $ m $ waveforms, each $ n $ 
samples long, $ \bH $ is a matrix with dimensions $ n \times m $. 
Using singular value decomposition~\cite{Strang}, $ \bH $ is 
factorized such that
\begin{equation}
  \bH = \bU \, \bS \, \bV ^\transpose,
\end{equation}
where $ \bU $ and $ \bV $ are orthonormal $ n \times r $ and
$ m \times r $ matrices, respectively, and $ \bS $ is a diagonal
$ r \times r $ matrix containing the singular values of $ \bH $
in decreasing order, i.e.\ $ \bS = \diag (s_1,...,s_r)$ with
$ s_1 \geq \ldots \geq s_r > 0 $,
$ r = \rank (\bH) \leq \min (m, n) $. The columns of $ \bU $ are the
eigenvectors of the empirical covariance matrix
$ \bH \, \bH^\transpose $, and similarly, the columns of $ \bV $ are
the eigenvectors of $ \bH^\transpose \bH $. Additionally, the
singular values in $ \bS $ are the square roots of the eigenvalues
$ \lambda_i $ of either $ \bH \, \bH^\transpose $ or
$ \bH^\transpose \bH $, i.e.\ $ s_i = \sqrt{\lambda_i} $.

The columns of $ \bU $, i.e.\ $ \bu_1, \ldots, \bu_r $, form an
orthonormal basis of the linear space spanned by the columns of
$ \bH $, i.e.\ the signal waveform space, and each signal waveform can
now be uniquely represented as a linear combination of these
eigenvectors. A measure of multivariate scatter about the mean is the
trace of the empirical covariance matrix,
$ \tr (\bH \, \bH^\transpose) $, also called \textsl{total variation},
which equals $ \sum_i^r \lambda_i $. So the sum of the first
$ k \leq r $ largest eigenvalues, $ \sum_{i = 1}^k \lambda_i $,
measures how much of the total variability of the waveforms is
explained by the first $ k $ eigenvectors. These are referred to as
the first $ k $ ``principal components'' (PCs) and achieve an optimal
dimension reduction from the $ r $-dimensional signal waveform space
to a $ k $-dimensional subspace.

Note that each waveform is $ n $ samples long and $ \bU $ has
dimensions $ n \times r $. For the waveforms considered in this
article, $ n $ is typically 1000 to 10\,000 samples long, so computing
the eigenvectors of $ \bH \, \bH^\transpose $ in $ \bU $ can be
computationally expensive. On the other hand, the number $ m $ of
waveforms in the catalog is of the order of 100. So, computing the
eigenvectors of $ \bH^\transpose \bH $ in $ \bV $ is much less demanding,
and they can be used to compute the eigenvectors in $ \bU $ by first
noting that
\begin{equation}
  \bH^\transpose \bH \bv_i = \lambda_i \bv_i,
\end{equation}
where $ \bv_i $ is the eigenvector of $ \bH^\transpose \bH $
corresponding to the $ i $th largest eigenvalue $ \lambda_i $. By
pre-multiplying both sides with $ \bH $, we obtain
\begin{equation}
  \bH \, \bH^\transpose \bH \bv_i = \lambda_i \bH \bv_i.
\end{equation}
From this, we can see that $ \bH \bv_i = \bu_i $, i.e.\ $ \bH \bv_i $
is the eigenvector corresponding to the $i$th largest eigenvalue of
$ \bH \, \bH^\transpose $ and thus equals the $ i $th column of
$ \bU $.
\begin{figure}
  \includegraphics[width=\columnwidth]{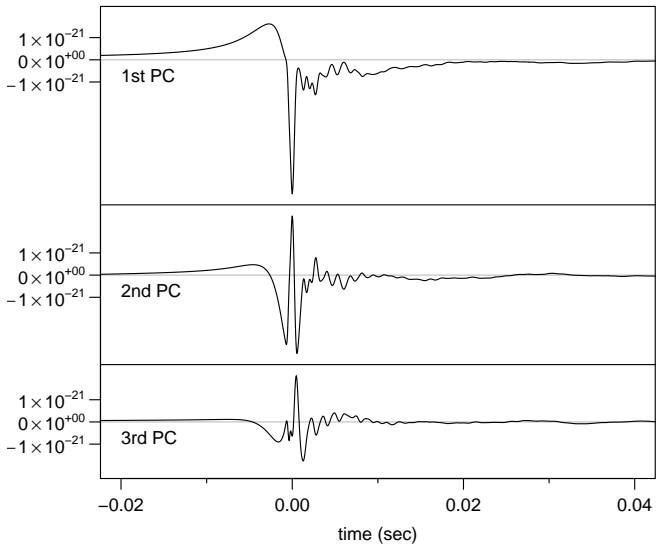}
  \caption{The top three principal components (PCs) 
           derived from the catalog of waveforms 
           described in Sec.~\ref{section:models}.}
  \label{fig:TopPCs}
\end{figure}
Fig.~\ref{fig:TopPCs} illustrates the resulting first three PCs
when applying this procedure to an actual waveform catalog.


\section{Analysis strategy}
\label{section:analysis}


\subsection{Definitions}
\label{subsection:definitions}

The starting point of our analysis is a waveform catalog, i.e.\ a
catalog of time series (of equal lengths) describing the gravitational
wave signal of core collapse events corresponding to different input
parameter settings. Let $ n $ denote the number of samples (discrete
time points) in each waveform vector and $ m $ the number of waveform
vectors in the catalog. As detailed in Section~\ref{section:pca}, for
these $ m $ time series we derive the first $ k < r $ eigenvectors,
$ \bx_1, \ldots, \bx_k $, corresponding to the $ k $ largest
eigenvalues~\cite{Heng2009}. Each of these eigenvectors again is 
of length~$ n $.

The data to be analyzed (the noisy measurement) are given in the form
of a time series vector $ \by $ of length $ N $ ($ N > n $). This
vector consists of additive non-white Gaussian noise with known
(one-sided) power spectral density denoted by $ S_1 (f) $, superimposed
by a core-bounce burst signal of length $ n $ located at an unknown
instant $ T $ along the time axis.

Our aim is to model the core-bounce burst signal observation in terms
of the basis of the $ k $ eigenvectors described above. To this end,
we assume that the mean of the $ y_i $'s is a linear combination
$ \beta_1 x_{i,1} + \ldots + \beta_k x_{i,k} $ of the $ k $
eigenvectors, and zero before and after the burst signal. In matrix
notation, this can be expressed in terms of the expectation value
$ \expect [\by] = \bX_{\!(T)} \bbeta $, where
$ \bbeta = (\beta_1, \ldots, \beta_k)^{\transpose} $ denotes the
vector of regression coefficients,  and the $ N \times k $ matrix
$ \bX_{\!(T)} $ has column vectors formed by the zero-padded $ k $
largest eigenvectors $ \bx_1, \ldots, \bx_k $ which are cyclically
time-shifted by a lag $ T $.

The signal reconstruction is eventually accomplished based on the
Fourier domain representations of data and signal; in the following we
will be referring to the conventions explicated in
Appendix~\ref{appendix:dft}. Let $ \tilde{\by} $ denote the Fourier
transformed data vector and $ \tilde{\bx}_i $, $ i = 1, \ldots, k $,
denote the discretely Fourier transformed eigenvectors after
zero-padding each to length $ N $. The real and imaginary parts of
these form the columns of the $ N \times k $ real-valued matrix
$ \tilde{\bX} $ (neglecting the redundant elements due to hermitian
symmetry).

One of the unknown parameters to be estimated is the signal's location
$ T $ along the time axis, and in order to match data and signal, one
needs to be able to shift both against each other in time. Let
$ \tilde{\bX}_{\!(T)} $ be the matrix of Fourier domain basis vectors
shifted in time so that these correspond to a particular signal arrival
time $ T $ (with respect to some pivotal time point). Time-shifting of
$ \tilde{\bX} $ by some lag $ T $ can be done directly in the frequency
domain by multiplying the original Fourier transform by a factor of
$ \exp (-2 \pi \imag f T) $.


\subsection{Model 1: Basic linear regression model}
\label{subsection:model_1}

If the data were an \textsl{exact} linear combination of first $ k $
principal components plus measurement noise, then the following
standard linear model would adequately model the situation:
\begin{equation}
  \label{eqn:TimeDomainModel}
  \by = \bX_{\!(T)} \,\bbeta + \bvarepsilon,
\end{equation}
where $ \bvarepsilon $ is the (Gaussian) noise vector with given
(one-sided) spectral density $ S_1 (f) $. In the frequency domain this
corresponds to
\begin{equation}
  \label{eqn:FourierDomainModel}
  \tilde{\by} = \tilde{\bX}_{\!(T)} \, \bbeta + \tilde{\bvarepsilon},
\end{equation}
where $ \tilde{\bvarepsilon} $ now is the Fourier transformed noise
vector. In Fourier domain, the real and imaginary components of the
noise vector $ \tilde{\bvarepsilon} $ then simply are independently
zero-mean Gaussian distributed with variances proportional to the
power spectral density $ S_1 (f) $~\cite{Finn1992,
  RoeverMeyerChristensen2008} (see also
Appendix~\ref{appendix:fourier_domain}).

What is known here are the data (measurement) $ \by $, the matrix
$ \bX $ of basis vectors (derived from the waveform catalog through
PCA), and the noise's power spectral density. The unknowns are the
coefficients $ \bbeta $ and the time parameter $ T $. The
\textsl{a~priori} information about these is expressed in the prior
distribution $ \prob(\bbeta, T) $, which is assumed to be uniform, i.e.\
any value is assumed equally likely.


\subsection{Model 2: Random effects regression model}
\label{subsection:model_2}

In general, the basis vectors will not allow one to
\textsl{completely} reconstruct the original signal, there will always
be a certain mismatch left unexplained when simplifying the problem to
the reduced set of PCs as in (\ref{eqn:FourierDomainModel})
above. Neglecting the mismatch term would result in overconfidence in
the reconstructed waveform and parameters. However, the simplified
model introduced above will still be of interest, as it provides good
approximations that are useful in the eventual implementation.

Adding an extra mismatch error component $ \bm $ to the model, a
\textsl{random effect} in statistical terminology, the model is now
\begin{equation}
  \label{eqn:model2}
  \tilde{\by} =
  \tilde{\bX}_{\!(T)} \, \bbeta + \tilde{\bm} + \tilde{\bvarepsilon}.
\end{equation}
In the following, we expect the original catalog to be sufficiently
densely populated, so that the mismatch is primarily due to the effect
of neglecting PCs, and not due to the signal being much unlike the
ones in the catalog. All we assume to be known about the mismatch is
that it contributes a certain fractional amount to the signal's
power. That amount was expressed in~\cite{Heng2009} in terms of the
\textsl{match parameter} $ \mu $. 
Taking the difference between
reconstructed and actual signal to be zero on average, and then
expecting a certain (fractional) power for the mismatch, 
we use a Gaussian distribution with corresponding mean and variance
for modelling the mismatch,
\begin{equation}
  \label{eqn:mismatchNormalDistn}
  \prob(m_i | \sigma_m^2) = \normaldistn (0, \sigma^2_m)
  \; \Leftrightarrow \;
  \prob(\tilde{m}_j | \sigma_m^2) =
  \normaldistn (0, {\textstyle \frac{N}{2}} \sigma^2_m),
\end{equation}
i.e.\ the mismatch $ \bm $ enters the model as an additional white
noise component with a one-sided power spectral density of
$ 2 \Delta_t \sigma_m^2 $.
The assumption of a Gaussian distribution 
is a simple and convenient choice here,
and according to maximum entropy theory it also 
constitutes the most conservative possible choice~\cite{Bretthorst1999}.
It also seems to perform well in practice, despite the fact that the
actual mismatch follows a rather heavy-tailed distribution 
with many near-zero and also a substantial number of extreme values.
Since the mismatch is supposed to scale
with the signal (i.e.\ the `relative mismatch' is assumed constant),
the mismatch's variance parameter $ \sigma_m^2 $ is set such that
\begin{equation}
  \label{eqn:sigmaApproxScale} \textstyle
  \sigma^2_m \approx \gamma^2 \, \frac{1}{N} \| \bX\bbeta \|^2
\end{equation}
depending on the PC coefficients $ \bbeta $ via the implied
sum-of-squares (or power) of the PC contribution to the signal,
\begin{equation}
  \frac{1}{N} \| \bX \bbeta \|^2 =
  \frac{1}{N} \sum_{i = 1}^N \Bigl( \sum_{j = 1}^Z \beta_j x_{i,j} \Bigr)^2,
\end{equation}
and the scaling factor $ \gamma^2 = \frac{1 - \mu^2}{\mu^2} $ is set
so that it corresponds to a particular match $ \mu $ as
in~\cite{Heng2009}. The above relationship between $ \mu $ and
$ \gamma^2 $ results from assuming the PC and mismatch contributions
to the signal $ \bs = \bX \bbeta + \bm $ to be (approximately)
orthogonal: $ \bm \perp \bX \bbeta $ (i.e.\ the mismatch is defined as
what is not spanned by the set of PCs).

The actual amount of mismatch is another unknown, and the
(approximate) scaling of $ \sigma^2_m $ with the signal power as in
Eq.~(\ref{eqn:sigmaApproxScale}) is ensured through the definition of
the prior distribution, which is set up as
\begin{equation}
  \prob (\bbeta, T, \sigma_m^2) =
  \prob (T) \times \!\!\!\!\!\! \underbrace{\prob (\bbeta, \sigma_m^2)}_{= \prob(\bbeta) \times \prob(\sigma_m^2 | \bbeta)}\!\!\!\!\!\!.
\end{equation}
As in the previous Section~\ref{subsection:model_1}, the priors
$ \prob (T) $ and $ \prob (\bbeta) $ are set to be independent and
uniform, but the conditional prior distribution
$ \prob (\sigma^2_m | \bbeta) $ is taken to be a \textsl{scaled
  inverse $ \chi^2 $ distribution}:
\begin{equation} \textstyle
  \prob_{\nu, \gamma^2} (\sigma_m^2 | \bbeta) =
  \invchisq \left( \nu, \,\gamma^2 \,\frac{1}{N} \| \bX\bbeta \|^2 \right)
\end{equation}
with \textsl{degrees-of-freedom} parameter $ \nu $ and \textsl{scale}
parameter $ (\gamma^2 \, \frac{1}{N} \| \bX\bbeta \|^2) $, so that the
prior certainty in the scale of $ \sigma^2_m | \bbeta $ is defined
through $ \nu $. For example, a specification of $ \gamma^2 = 0.1 $
and $ \nu = 3 $ implies for the prior that
$ \prob \bigl( 0.038 < \sigma^2 / (\frac{1}{N} \| \bX\bbeta \|^2) < 0.85 \bigr) 
= 90\% = \prob (0.73 < \mu < 0.98) $, and a conditional prior mean of
$ \expect [\sigma_m^2 | \bbeta] =
\frac{\nu}{\nu-2} \gamma^2 \, \frac{1}{N} \| \bX\bbeta \|^2 $. Setting
$ \nu = 0 $ yields the (improper) Jeffreys prior, which does not
depend on its prior scale parameter~\cite{BDA}. The $ \invchisq $
distribution was chosen here because it constitutes the
\textsl{conjugate prior distribution} for this problem~\cite{BDA},
which makes it a `natural' choice and makes the eventual
implementation particularly simple, as will be seen in
Section~\ref{subsection:mcmc} below. The parameters $ \gamma^2 $ and $
\nu $ may now be set so that the prior distribution reflects the
reconstruction accuracy to be expected from the given set of $ k $
basis vectors derived from the waveform catalog at hand.


\subsection{Monte Carlo integration}
\label{subsection:mcmc}

Inference on waveforms and parameters usually requires integrating the
parameters' posterior distribution, as one is interested in figures
like posterior expectations, quantiles, or marginal distributions.
These are here determined using stochastic (Monte Carlo) integration,
i.e.\ by generating samples from the posterior probability
distribution and then approximating the desired integrals by sample
statistics (means by averages, etc). The generation of samples from
the posterior distribution is done using Markov chain Monte Carlo (MCMC) 
methods, that is, by designing a Markov process whose stationary
distribution is the posterior probability distribution of interest,
and which may then be numerically simulated step-by-step to produce
the desired posterior samples. The generated samples then allow one to
explore marginal or joint distributions of individual parameters, or
of functions of the parameters as in the case of signal
reconstruction, where the posterior distribution of
$ \tilde{\bX}_{\!(T)} \bbeta + \tilde{\bm} $ is of interest, and
$ T $, $ \bbeta $ and $ \tilde{\bm} $ are random variables.

In the case of the basic linear model (see
Section~\ref{subsection:model_1}), the set of unknown parameters
consists of the vector of PC coefficients $ \bbeta $ and the time
shift parameter $ T $. An MCMC algorithm here may be implemented as a
\textsl{Gibbs sampler}~\cite{BDA}, since the conditional
posterior distribution $ \prob (\bbeta|T, \by) $ of coefficients
$ \bbeta $ for a given time $ T $ is already known and easy to sample
from (see Eq.~(\ref{eqn:conditionalNormalPosterior}) below). In a
Gibbs sampler, generating samples from the joint distribution is
carried out by alternately sampling from the two conditional
distributions $ \prob (\bbeta | T, \by) $ and
$ \prob (T | \bbeta, \by) $~\cite{BDA}. Samples from
$ \prob (\bbeta | T, \by) $ can be generated straight away, while a
simple ``Metropolis-within-Gibbs'' step, i.e.\ a nested Metropolis
sampler implementation~\cite{BDA}, may be utilized for sampling from
from the (one-dimensional) distribution $ \prob (T | \bbeta, \by) $.

For the basic linear regression model (see
Section~\ref{subsection:model_1}), the conditional posterior
distribution of the PC coefficients $ \bbeta $ for a given time shift
$ T $ is a multivariate Gaussian distribution:
\begin{equation}
  \label{eqn:conditionalNormalPosterior}
  \prob(\bbeta | T, \by) = \normaldistn (\hat{\bmu}_T, \hat{\bSigma}_T),
\end{equation}
where
\begin{equationarray}
  \hat{\bSigma}_T & = &
  (\tilde{\bX}_{\!(T)}^\transpose D^{-1} \tilde{\bX}_{\!(T)})^{-1},
  \\
  \hat{\bmu}_T & = & (\tilde{\bX}_{\!(T)}^\transpose D^{-1}
  \tilde{\bX}_{\!(T)})^{-1} \tilde{\bX}_{\!(T)}^\transpose D^{-1} \tilde{\by},
\end{equationarray}%
and $ D = \diag \bigl( \sigma^2 (f_j) \bigr) $ is the noise's
covariance matrix, a diagonal matrix of the individual variances, as
given in Eq.~(\ref{eqn:FDomainGaussian})~\cite{BDA}.

Sampling from the posterior distribution works similarly for the
extended model (see Section~\ref{subsection:model_2}), in addition to
$ \bbeta $ and $ T $, the mismatch $ \tilde{\bm} $ and mismatch
variance $ \sigma_m^2 $ need to be sampled from. Note that the
conditional posterior distribution
$ \prob (\bbeta | T, \tilde{\bm}, \sigma_m^2, \by) $ is not simply
multivariate Gaussian any more, since changes in $ \bbeta $ lead to
different prior density values (via its effect on the signal's
sum-of-squares value). The expression in
Eq.~(\ref{eqn:conditionalNormalPosterior}) still is an excellent
approximation and is useful for defining a proposal distribution
within the Metropolis step of the algorithm. The conditional
distribution of the mismatch vector $ \tilde{\bm} $ is independent
Gaussian:
\begin{equation}
  \prob(\tilde{\bm} | \bbeta, T, \sigma_m^2, \by) =
  \normaldistn (\check{\bmu}, \check{\bSigma}),
\end{equation}
where
\begin{equationarray}
  \check{\mu}_{j, \re} & = &
  \re ((\tilde{\by} - \tilde{\bX}_{\!(T)} \, \bbeta)_j) \,
  \frac{\frac{N}{2} \sigma_m^2}{\frac{N}{2} \sigma_m^2 + \frac{N}{4\Delta_t} S_1 (f_j)},
  \\
  \check{\sigma}^2_{jj, \re} & = &
  \frac{\frac{N}{2} \sigma_m^2 \, \frac{N}{4 \Delta_t}
  S_1 (f_j)}{\frac{N}{2} \sigma_m^2 + \frac{N}{4 \Delta_t} S_1 (f_j)},
\end{equationarray}%
and analogously for the imaginary parts $ \im (\tilde{m}_j) $. For the
mismatch variance $ \sigma_m^2 $ the conditional posterior is
\begin{equation}
  \prob(\sigma_m^2 | \by, \bbeta, T, \tilde{\bm}) = \invchisq (\kappa, s^2),
\end{equation}
where the degrees of freedom $ \kappa $ and scale $ s^2 $ are
\begin{equationarray}
  \kappa & = & \nu + N ,
  \\
  s^2 & = &
  \frac{\nu\, \gamma^2\, {\textstyle\frac{1}{N} \| \bX\bbeta \|^2} + \sum_{i = 0}^N m_i^2}{\nu + N}
\end{equationarray}%
(see also Appendix~\ref{appendix:posteriors}).

The MCMC sampler was eventually implemented as a Gibbs sampler,
alternately sampling from the three conditional distributions of
\begin{enumerate}
\item $ \quad \bbeta, T  | \by, \tilde{\bm}, \sigma_m^2 $,
\item $ \quad \tilde{\bm} | \by, \bbeta, T, \sigma_m^2 $, and
\item $ \quad \sigma_m^2 | \by, \bbeta, T, \tilde{\bm} $.
\end{enumerate}
The first step is done in a ``Metropolis-within-Gibbs'' step, using a
symmetric proposal distribution for $ T $, the approximated posterior
of $ \bbeta | T, \ldots $ from
Eq.~(\ref{eqn:conditionalNormalPosterior}) for the corresponding
$ \bbeta $ proposal, and accepting/rejecting as in a usual
Metropolis--Hastings sampler based on corresponding posterior and
proposal probability density values. Samples for the second and third step
may be generated directly.


\section{Implementation and application}
\label{section:implementation}


\subsection{Setup}
\label{subsection:setup}

In the following examples we are using the waveform catalog described
in Section~\ref{section:models}, containing 128 gravitational
radiation waveforms from rotating core collapse and
bounce~\cite{DimmelmeierEtAl2008a, DimmelmeierWaveCatalogue}. The
basis vectors to be used for signal reconstruction are generated
through a PCA (see Section~\ref{section:pca})~\cite{Heng2009}.
Utilizing the same code and general setup for calculating the models
described in~\cite{DimmelmeierEtAl2008a, DimmelmeierWaveCatalogue}, we
then compute three new rotating stellar core collapse models with input
parameter values that did not appear in the original catalog (but lie
within the range of the catalog parameter space) and their associated
waveforms. The aim is then to reconstruct these waveforms
using the PCA method.

The simulated data used here is $ 1 \mathrm{\ s} $ in length and sampled at
$ 16\,384 \mathrm{\ Hz} $, superimposed with (simulated) non-white
Gaussian noise, and Tukey-windowed. The shape of the noise curve is
here taken to correspond to LIGO at its initial sensitivity as stated
in~\cite{DamourIyerSathyaprakash2001}; this is the definition that is
also implemented in the 
LIGO scientific collaboration's
LSC algorithm library (LAL)~\cite{LAL-Manual}.
The noise's spectral density $ S_1 (f) $ is estimated by averaging
over a thousand ``empirical'' periodograms of identically generated
data (same size and resolution, same windowing applied), more closely
resembling a realistic case in which the noise spectrum would need to
be estimated as well. In addition,  this way convolution effects on
the spectrum due to the finite data size and windowing are
compensated.

\begin{figure}
  \includegraphics[width=\columnwidth]{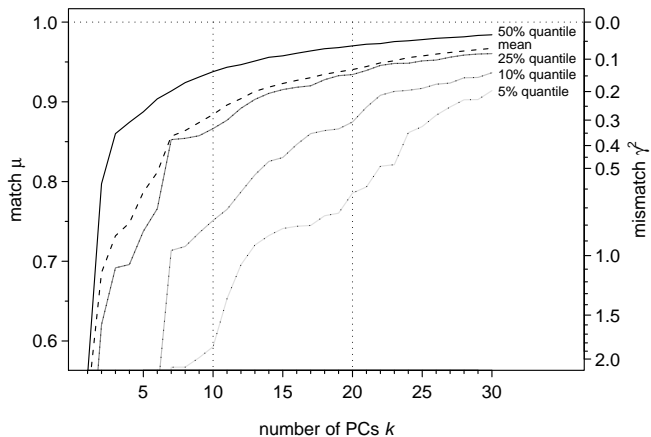}
  \caption{Achievable match $ \mu $ and scaling factor $ \gamma^2 $
    for given numbers $ k $ of PCs across the whole catalog used
    here. A particular number $ k $ of PC basis vectors yields a
    certain match $ \mu $ for each of the 128 waveforms in the
    catalog. Shown here is the distribution of matches across the
    catalog for increasing values of $ k $, as characterised by mean,
    median and some more quantiles.}
  \label{fig:CatalogMismatch}
\end{figure}

\begin{figure*}
  \includegraphics[width=\textwidth]{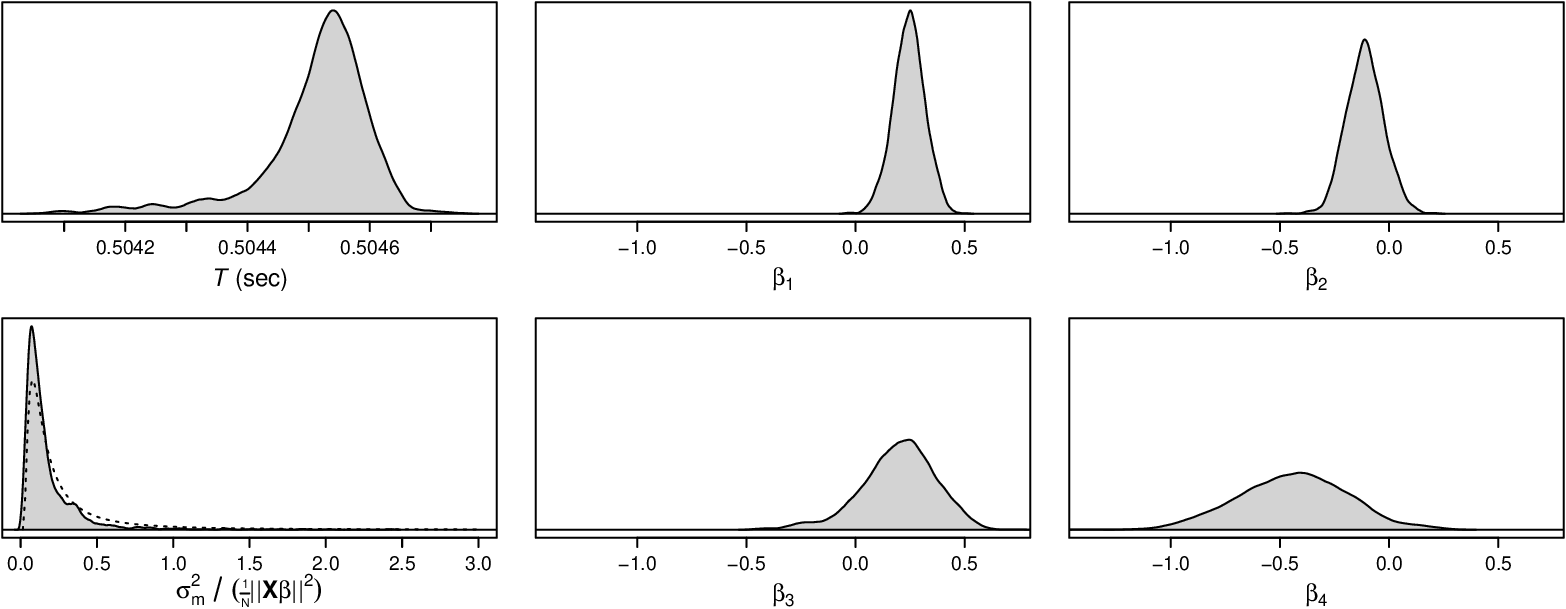}
  \caption{Marginal posterior probability distributions of the parameters of
    example signal {\exA} with SNR $ \rho = 10 $,
    indicating the values and uncertainties of the model parameters 
    as inferred from the data. 
    Only the first four
    coefficients $ \beta_1 $ to $ \beta_4 $ corresponding 
    to the top four principal components are shown here; there are
    $ k =10 $ coefficients in total. 
    The dashed line in the top right plot shows the prior probability distribution 
    in comparison to the (very similar) posterior.}
  \label{fig:Marginals1}
\end{figure*}

The prior distribution for the mismatch parameter $ \sigma_m^2 $ is
set by determining what match $ \mu $ would be achievable for the
given number $ k $ of PCs. This is done by projecting each single
waveform in the catalog onto the span of the PCs, and then (for any
given number $ k $ of PCs) considering the distribution of achieved
mismatch values across the catalog. Fig.~\ref{fig:CatalogMismatch}
illustrates the mean, median and some more quantiles of the mismatch
distribution for increasing numbers of PCs (in analogy to Fig.~2
in Ref.~\cite{Heng2009}). We chose the prior's parameters (scale
$ \gamma^2 $ and degrees-of-freedom $ \nu $) so that it matched the
distribution of computed matches across the catalog. For the two
example settings of $ k = 10 $ and $ k = 20 $ considered in the
following, this leads to maximum likelihood estimates of
($ \gamma^2 = 13.3\% $, $ \nu = 2.82 $) and ($ \gamma^2 = 5.70\% $,
$ \nu = 2.96 $). The resulting prior density for $ k = 10 $ PCs is
also shown in Fig.~\ref{fig:Marginals1}.


\subsection{Examples}
\label{subsection:examples}

The first example illustrated here, labeled signal {\exA} in the
following, has a signal-to-noise ratio (SNR) of $ \rho = 10 $,
where the SNR is defined as
\[\rho = \sqrt{4\sum_j\frac{\frac{\Delta_t}{N} |\tilde{h}(f_j)|^2}{S_1(f_j)}}.\]
It was generated assuming the \textsl{Shen} EoS, a mass
$ M_\mathrm{prog} = 20 \, \solarmass $ for the progenitor, a
precollapse central angular velocity
$ \Omega_\mathrm{c,i} = 5.48 \mathrm{\ rad\ s}^{-1} $, a
differentiality of the precollapse rotation profile given by
$ A = 1000 \mathrm{\ km}$, and an effective distance of
$ d_\mathrm{e} = 5.17 \mathrm{\ kpc} $. The effective distance depends
on the actual distance to the source, and also reflects the effect on
the amplitude of the gravitational wave signal from other parameters
(such as the geometry of the detector with respect to the source), and
is in general greater than the actual distance. 
The signal is embedded
within simulated interferometer noise, and, using the model with $ k =
10 $ basis vectors, the posterior distributions of the model
parameters are derived using an MCMC implementation as described in
the previous Section~\ref{subsection:mcmc}.
The maximum achievable match for this example waveform 
(for $k=10$ and varying time shift~$T$) is $\mu=0.97$.

\begin{figure}
  \includegraphics[width=\columnwidth]{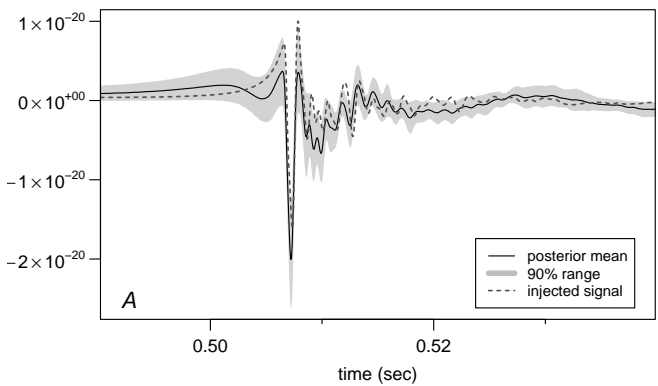}
  \caption{Reconstruction of the example signal {\exA} with SNR
    $ \rho = 10 $ using $ k = 10 $ PCs.}
  \label{fig:Reconstruction1}
\end{figure}

The marginal posterior distributions of some individual parameters are
illustrated in Fig.~\ref{fig:Marginals1}. Note the timing accuracy,
which has a standard error of $ 0.1 \mathrm{\ ms} $ in this case; in
the following examples (for the same SNR), these are also of the order
of sub-millisecond. The parameters' joint posterior distribution
provides the posterior probability distribution of the waveform at any
given time $ t_i $. Fig.~\ref{fig:Reconstruction1} shows the posterior
mean and a 90\% confidence band in comparison with the originally
injected waveform.

\begin{figure}
  \includegraphics[width=\columnwidth]{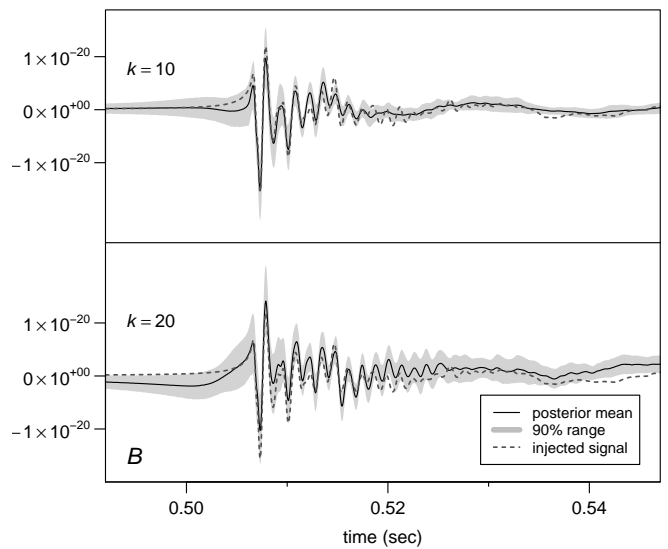}
  \caption{A repeated analysis and reconstruction of the same example
    signal {\exB} with SNR $ \rho = 10 $ using different numbers of
    PCs ($ k = 10 $ and $ k = 20 $). Note that a larger number of
    basis vectors in the model does not necessarily improve signal
    recovery.}
  \label{fig:Reconstruction2}
\end{figure}

The example signal {\exB} illustrates how the signal recovery changes
with differing numbers of PCs included in the model. This signal was
generated assuming the \textsl{LS} EoS, a mass
$ M_\mathrm{prog} = 15.0 \, \solarmass $ for the progenitor, a
precollapse central angular velocity
$ \Omega_\mathrm{c,i} = 2.825 \mathrm{\ rad\ s}^{-1} $, a
differentiality of the precollapse rotation profile of
$ A = 1000 \mathrm{\ km} $, and an effective distance of
$ d_\mathrm{e} = 3.24 \mathrm{\ kpc} $. 
The signal's match is $\mu=0.89$ when using 10~PCs, 
and $\mu=0.95$ for 20~PCs.
The same data are analyzed
twice, once using $ k = 10 $ and then using $ k = 20 $ PCs; the
resulting recovered waveforms are shown in
Fig.~\ref{fig:Reconstruction2}. The enlarged model in principle allows
for a better match of the signal, but on the other hand the larger
number of parameters also decreases the certainty in parameter
estimates, so that the recovery does not necessarily improve.
In this example, the posterior variances of the PC parameters common
to \textsl{both} models ($ \beta_1, \ldots, \beta_{10}$) as well as
the time parameter $ T $ increased for the $ k = 20 $ case. In the
resulting signal reconstruction (see Fig.~\ref{fig:Reconstruction2})
the confidence band is wider, and the discrepancy between injected
signal and posterior mean also increases.

\begin{figure}
  \includegraphics[width=\columnwidth]{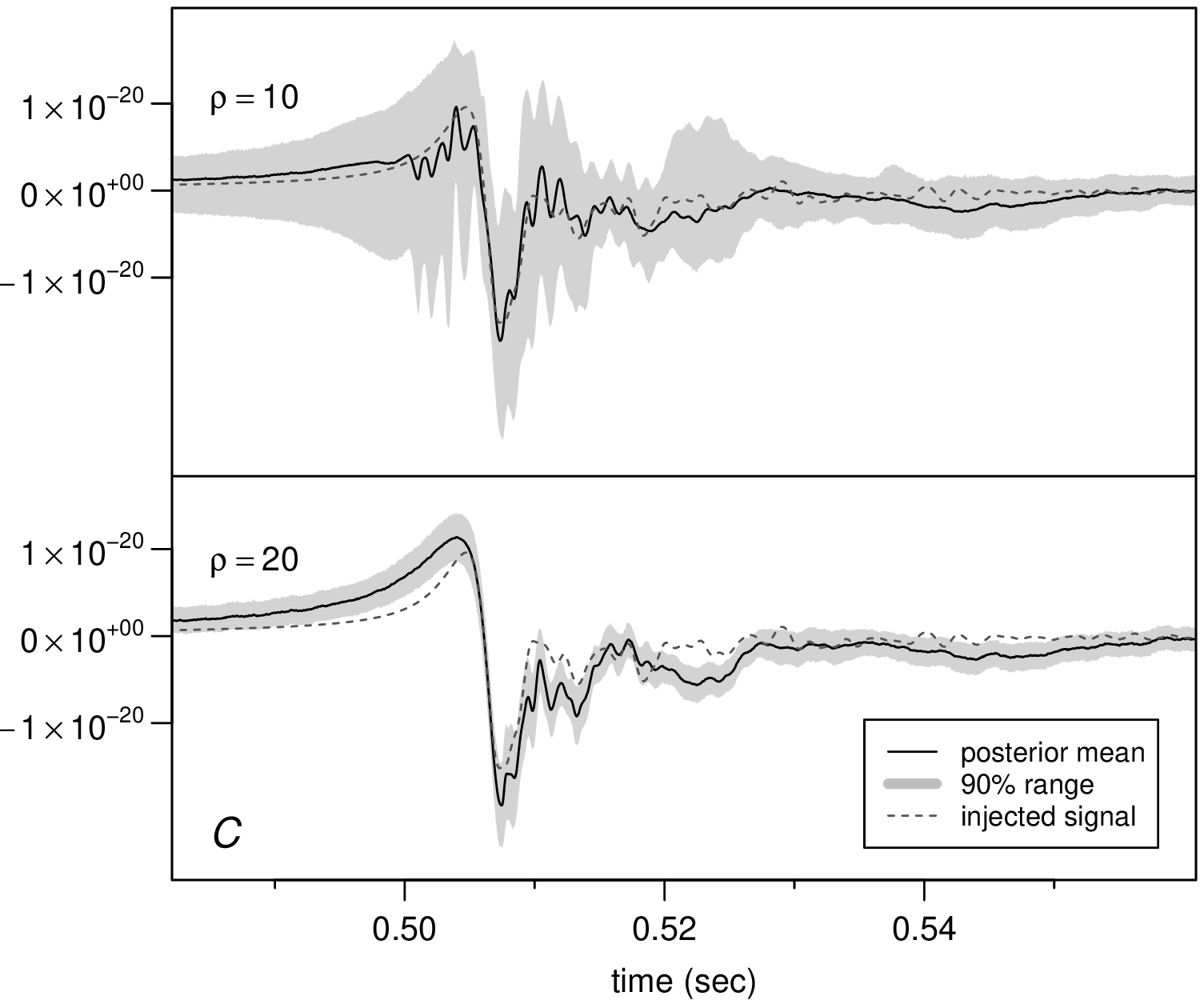}
  \caption{Reconstruction of the same example signal {\exC} at
    different overall amplitudes (and, with that, SNRs: $ \rho = 10 $
    and $ \rho = 20 $), both times using $ k = 10 $ PCs.}
  \label{fig:Reconstruction3}
\end{figure}

The reconstruction of signals at different SNRs can be seen in the
following example signal {\exC} in Fig.~\ref{fig:Reconstruction3};
here the same signal is injected at different overall amplitudes and
hence SNRs ($ \rho = 10 $ and $ \rho = 20 $). The injected signal was
generated using the LS EoS, a mass
$ M_\mathrm{prog} = 40.0 \, \solarmass $ for the progenitor, a
precollapse central angular velocity
$ \Omega_\mathrm{c,i} = 9.596 \mathrm{\ rad\ s}^{-1} $, a
differentiality of the precollapse rotation profile given by
$ A = 500 \mathrm{\ km} $, and effective distances of
$ d_\mathrm{e} = 8.32 $ and $ 4.16 \mathrm{\ kpc} $, respectively.
The resulting signal waveform's match is $\mu=0.97$.
As expected, a higher SNR yields a more accurate recovery, and the
resulting posterior variances are correspondingly smaller.


\section{Correspondence between principal components and physical
  parameters}
\label{section:correspondence}

The analysis performed above not only allows one to reconstruct the
waveform, but the posterior distribution of the PC coefficients
actually also contains information on the signal's physical
parameters ($ \Omega_\mathrm{c,i} $, $ A $, $ M_\mathrm{prog} $, and
EoS) as well as other evolution parameters (like
$ \rho_\mathrm{max,b} $, $ T_\mathrm{b} / |W|_\mathrm{b} $, and
$ T_\mathrm{pb} / |W|_\mathrm{pb} $). In the present study we are only
able to choose between two specific EoSs, but as this work progresses
the goal is to make estimates from a wider selection of EoSs.
Considering the catalog of waveforms alone, one can project each
waveform onto the span of the PCs (as was done in the construction of
Fig.~\ref{fig:CatalogMismatch}) and inspect the resulting fitted PC
coefficients.
\begin{figure}
  \includegraphics[width=0.75\columnwidth]{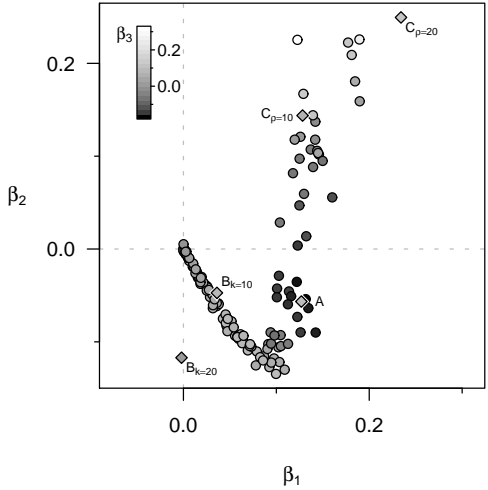}
  \caption{The resulting first three PC coefficients $ \beta_1 $,
    $ \beta_2 $, and $ \beta_3 $ when matching each of the 128
    waveforms in the catalog against the set of PCs (no noise
    involved). The waveforms in the catalog all correspond to a signal
    from a fixed distance ($ 10 \mathrm{\ kpc} $). Note that the
    coefficients only occupy a very restricted region in parameter
    space. The diamonds show the posterior mean values of the PC
    coefficients for the three examples (labeled A, B, and C)
    discussed in the text. For comparison they are scaled down (by
    their known distance) to the same magnitude.}
  \label{fig:CatalogCoefs1}
\end{figure}
Fig.~\ref{fig:CatalogCoefs1} illustrates the values of the first three
coefficients ($ \beta_1 $, $ \beta_2 $, and $ \beta_3 $) for all the
signals in the catalog. Obviously the distribution of coefficients in
parameter space exhibits some structure. This structure can now be
exploited, the idea being that the posterior means and variances of
the coefficients $ \beta_i $ for a measured signal expose a pattern
that is characteristic for the type of signal. For illustration, in
the following we describe a simple approach involving the three
example waveforms from the previous section. The coefficients'
posterior distribution can be related back to the sets of coefficients
for reconstructing every signal in the original catalog, in order to
identify similarities. A simple \textsl{ad hoc} measure of similarity
is a $ \chi^2 $ match that relates a set of best-matching coefficients
$ \hat{\beta}_i $ ($ i = 1, \ldots, k $) associated with each catalog
entry to the posterior distribution of the coefficients
$ \prob (\beta_i | y) $ of the inferred signal. Such a match may be
defined as
\begin{equation}
  \label{eqn:ChisqMatch}
  \chi^2 = \min_{c \in \realline} \sum_{i = 1}^k
  \frac{(c \hat{\beta}_i - \expect[\beta_i | \by])^2}{\var (\beta_i | \by)},
\end{equation}
considering only posterior means and variances of individual
coefficients $ \beta_i $, and allowing for a scaling factor $ c $ in
the overall amplitude of the matched signal. Computing the match for
each signal from the catalog allows one to rank all signals and then
determine those that fit the best. A mapping scheme set up this way
combines elements of \textsl{nearest-neighbor} and
\textsl{naive-Bayes} classification
techniques~\cite{HastieTibshiraniFriedman}.

\begin{figure}
  \includegraphics[width=\columnwidth]{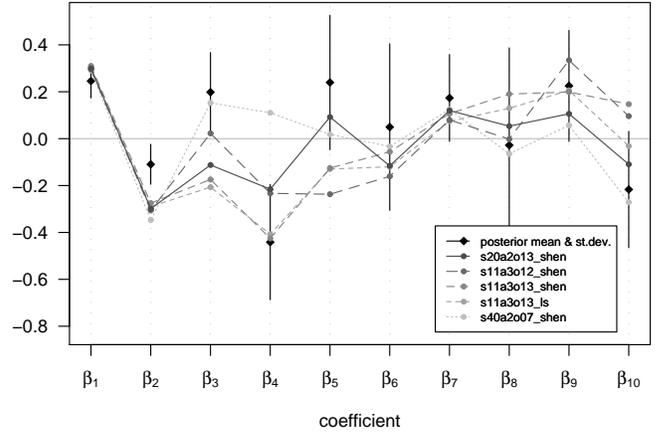}
  \caption{Comparison of the distributions of all PC coefficients $
    \beta_i $ to the best-matched sets of coefficients found in the
    original catalog. Posterior means and standard deviations (`error
    bars') are shown in black, the (scaled) sets of the top~five
    best-matching coefficients are shown in different shades of gray.}
  \label{fig:ChisquaredMatch}
\end{figure}

For the initial example signal {\exA} shown above (see
Figs.~\ref{fig:Marginals1} and~\ref{fig:Reconstruction1}), the
posterior means and standard deviations (`error bars') are illustrated
in a parallel coordinates plot~\cite{Inselberg1985} in
Fig.~\ref{fig:ChisquaredMatch}, together with the five best-matching
catalog entries. More details on the injected and the ten
best-matching signals are given in Table~\ref{tab:bestMatches}.
Table~\ref{tab:bestMatches2} lists the closest correspondents for the
other example signals (examples {\exB} and {\exC}) in comparison with
the injection values. In the case of signal~{\exA}, the best matching
waveform from the catalog and the injected signal have very similar
physical parameters. For signals~{\exB} and {\exC} this simple method
still provides a good  estimation of the physical parameters, although
admittedly not as accurate as case~{\exA}; 
note that the reconstruction of the waveform is quite accurate for all three signals. 
We think that through the use of techniques such as
Procrustes rotation~\cite{MardiaKentBibby} we will be able to provide
a more direct link to the physical parameters. The posterior mean
values for the first three PC coefficients ($ \beta_1 $ to
$ \beta_3 $) for all five examples are also illustrated in
Fig.~\ref{fig:CatalogCoefs1}. Reconstruction of the measured signals
through the PC basis vectors not only allows one to capture the the
waveforms' appearances, but also yields information on the underlying
physical parameters.

\begin{table*}[ht]
  \caption{The injected example signal {\exA} and the
    corresponding top 10 best matching catalog entries. The column
    labeled with `model' identifies the signal from the original
    catalog~\cite{DimmelmeierWaveCatalogue}. The following columns
    state achieved $ \chi^2 $ match (according to
    Eq.~(\ref{eqn:ChisqMatch})), the corresponding EoS, mass of the
    progenitor model ($ M_\mathrm{prog} $), precollapse angular velocity at the
    center ($ \Omega_\mathrm{c,i} $), precollapse differential
    rotation length scale ($ A $), 
    rotation rates initially ($ T_\mathrm{i} / |W|_\mathrm{i} $), 
    at the time of bounce ($ T_\mathrm{b} / |W|_\mathrm{b} $), 
    and late after bounce ($ T_\mathrm{pb} / |W|_\mathrm{pb} $), 
    and the maximum density in the core at the time of bounce 
    ($ \rho_\mathrm{max,b} $)~\cite{DimmelmeierEtAl2008a}.
    The corresponding best-matching effective distance
    $ d_\mathrm{e} $ results from inverting the amplitude $ c $ (see
    Eq.~(\ref{eqn:ChisqMatch})) that yields the optimal match.}
  \label{tab:bestMatches}
  \begin{ruledtabular}
    \begin{tabular}{clcccccccc}
      rank &
      \multicolumn{1}{c}{model} &
      $ \chi^2 $ &
      EoS &
      $ M_\mathrm{prog} $ &
      $ \Omega_\mathrm{c,i} $ &
      $ A $ &
      $ \frac{T_\mathrm{i}}{|W|_\mathrm{i}} \,\Big|\, \frac{T_\mathrm{b}}{|W|_\mathrm{b}} \,\Big|\, \frac{T_\mathrm{pb}}{|W|_\mathrm{pb}} $ &
      $ \rho_\mathrm{max,b} $ &
      $ d_\mathrm{e} $ \\
      &
      &
      &
      &
      [$ \solarmass $] &
      [$ \mathrm{rad\ s}^{-1} $] &
      [km] &
      [\%] &
      [$ 10^{14} \mathrm{\ g\ cm}^{-3} $] &
      [kpc] \\
      \hline
      \multicolumn{3}{r}{\textsl{example signal {\exA} (injected)}} & Shen & $20.0$ & $\pz 5.48$ & $1000$ & $1.27\,|\,11.9\,|\,10.5$ & $2.62$ & $5.17$ \\
      \hline
      $ \pz 1$ & \texttt{s20a2o13\_shen} & $10.8$ & Shen & $20.0$ & $\pz 6.45$ &    $1000$ & $1.80\,|\,14.8\,|\,   12.8$ & $2.42$ & $3.50$ \\
      $ \pz 2$ & \texttt{s11a3o12\_shen} & $12.8$ & Shen & $11.2$ &    $10.65$ & $\pz 500$ & $1.28\,|\,14.9\,|\,   12.3$ & $2.46$ & $3.46$ \\
      $ \pz 3$ & \texttt{s11a3o13\_shen} & $13.7$ & Shen & $11.2$ &    $11.30$ & $\pz 500$ & $1.44\,|\,16.1\,|\,   13.2$ & $2.36$ & $3.37$ \\
      $ \pz 4$ & \texttt{s11a3o13\_ls}   & $13.8$ & LS   & $11.2$ & $   11.30$ & $\pz 500$ & $1.44\,|\,15.8\,|\,   13.2$ & $2.64$ & $3.10$ \\
      $ \pz 5$ & \texttt{s40a2o07\_shen} & $14.6$ & Shen & $40.0$ & $\pz 3.40$ &    $1000$ & $0.72\,|\,11.8\,|\,\pz 9.9$ & $2.58$ & $3.75$ \\
      $ \pz 6$ & \texttt{s11a3o12\_ls}   & $15.4$ & LS   & $11.2$ & $   10.65$ & $\pz 500$ & $1.28\,|\,14.7\,|\,   12.3$ & $2.75$ & $3.12$ \\
      $ \pz 7$ & \texttt{s15a2o09\_shen} & $15.4$ & Shen & $15.0$ & $\pz 4.56$ &    $1000$ & $1.09\,|\,11.8\,|\,   10.3$ & $2.58$ & $3.42$ \\
      $ \pz 8$ & \texttt{s20a3o09\_shen} & $15.7$ & Shen & $20.0$ & $\pz 8.99$ & $\pz 500$ & $0.90\,|\,15.7\,|\,   12.5$ & $2.38$ & $3.58$ \\
      $ \pz 9$ & \texttt{s20a3o07\_shen} & $17.4$ & Shen & $20.0$ & $\pz 5.95$ & $\pz 500$ & $0.50\,|\,10.1\,|\,\pz 8.0$ & $2.70$ & $3.24$ \\
      $    10$ & \texttt{s20a2o13\_ls}   & $20.6$ & LS   & $20.0$ & $\pz 6.45$ &    $1000$ & $1.80\,|\,14.4\,|\,   12.9$ & $2.75$ & $2.94$ \\
      \vdots & \multicolumn{1}{c}{\vdots} & \vdots & \vdots & \vdots & \vdots & \vdots & \vdots & \vdots  & \vdots \\ [0.5ex]
    \end{tabular}
  \end{ruledtabular}
\end{table*}

\begin{table*}[ht]
  \caption{The injected signal example signals {\exB} and {\exC}, and
    the corresponding best matching catalog entries when varying
    $ k $ or $ \rho $, respectively; see also
    Section~\ref{subsection:examples} and
    Table~\ref{tab:bestMatches}.}
  \label{tab:bestMatches2}
  \begin{ruledtabular}
    \begin{tabular}{clcccccccc}
      PCs / SNR &
      \multicolumn{1}{c}{model} &
      $ \chi^2 $ &
      EoS &
      $ M_\mathrm{prog} $ &
      $ \Omega_\mathrm{c,i} $ &
      $ A $ &
      $ \frac{T_\mathrm{i}}{|W|_\mathrm{i}} \,\Big|\, \frac{T_\mathrm{b}}{|W|_\mathrm{b}} \,\Big|\, \frac{T_\mathrm{pb}}{|W|_\mathrm{pb}} $ &
      $ \rho_\mathrm{max,b} $ &
      $ d_\mathrm{e} $ \\
      &
      &
      &
      &
      [$ \solarmass $] &
      [$ \mathrm{rad\ s}^{-1} $] &
      [km] &
      [\%] &
      [$ 10^{14} \mathrm{\ g\ cm}^{-3} $] &
      [kpc] \\
      \hline
      \multicolumn{3}{r}{\textsl{example signal {\exB} (injected)}} & LS & $15.0$ & $\pz 2.82$ & $\pz\, 1000$ & $0.41\,|\,\pz 5.5\,|\,\pz 5.2$ & $3.61$ & $3.24$\\
     $k=10$ & \texttt{s11a3o07\_shen} & $\pz 1.9$ & Shen & $20.0$ & $\pz 5.95$ &    $\pz\,\pz 500$ & $0.40\,|\,\pz 5.9\,|\,\pz 5.1$ & $2.89$ & $3.41$\\
     $k=20$ & \texttt{s15a1o13\_ls} & $47.4$ & LS & $15.0$ & $\pz 2.71$ &    $50\,000$ & $3.26\,|\,\pz 6.1\,|\,\pz 6.6$ & $3.56$ & $1.28$\\
  \hline
     \multicolumn{3}{r}{\textsl{example signal {\exC} (injected)}} & LS & $40.0$ & $\pz 9.60$ & $\pz\,\pz 500$ & $1.46\,|\,22.1\,|\,19.0$ & $0.73$ & $8.32\,|\,4.16$\\
     $\rho=10$ & \texttt{s40a3o13\_ls}   & $\pz 5.8$ & LS  & $40.0$ & $   11.30$ &    $\pz\,\pz 500$ & $2.07\,|\,23.4\,|\,20.7$ & $0.28$ & $5.38$\\
     $\rho=20$ & \texttt{s40a2o15\_shen} &    $19.6$ & Shen & $40.0$ & $\pz 7.60$ &    $\pz\, 1000$ & $3.62\,|\,21.1\,|\,20.3$ & $0.27$ & $2.32$\\
   \end{tabular}
 \end{ruledtabular}
\end{table*}

The information available from matching a signal candidate against a
set of basis vectors may not only facilitate a classification within a
set of possible astrophysical sources (as illustrated in
Fig.~\ref{fig:CatalogCoefs1}), but may also help in distinguishing it
from signals of different origin, like instrumental glitches,
etc. Different types of potential signals may also be partly
reconstructible in terms of a linear combination of PCs, but a poor
match, an unusual combination of coefficients, or a better match from
an alternative set of basis vectors may then provide evidence in favor
or against certain signal origins.


\section{Conclusions and outlook}
\label{section:conclusions}

In this paper we have described initial work on implementing a scheme
by which the physical parameters associated with a rotating collapse
and bounce in a stellar core collapse event can be estimated through
the observation of gravitational waves by detectors such as LIGO,
VIRGO, GEO or TAMA. We have presented a technique which allows one to
reconstruct the detected signal through the use of numerically
calculated gravitational radiation waveforms, principal component
analysis, and Markov chain Monte Carlo (MCMC) methods; we have then
compared the reconstructed signals to the table of numerically
calculated waveforms, and inferred the physical parameters.

The results displayed in this paper were achieved through the application of
a number of different and difficult analysis techniques.
We are encouraged that in this initial study we have displayed that
these methods can be combined in a way such that physical information about
the supernova event can be extracted from the detected gravitational wave
data.  We have found that our method is quite successful in the
reconstruction of the detected waveform, and this is displayed in the three
examples. In the present analysis we have used the output from a single
detector, and we expect the accuracy of the signal reconstruction will only
get better as we expand the application of this work to the coherent
analysis of multiple interferometers. We were able to make a fairly good
association with the physical parameters; however, we think that we will be
able to greatly improve on our ability to make parameter estimation
estimates. In the present study we worked with a relatively small signal
catalog, where each physical parameter took on only a handful of different
values; for example, there were only two EoS to choose from. Our parameter
estimation demonstration was pretty good within these constraints, but we
expect the accuracy to increase with expanded signal catalogs. In addition,
Procrustes rotation~\cite{MardiaKentBibby} is an example of one technique
that we intend to apply in order to make better statistical estimates of the
physical parameters based on the eigenstates provided by the PCA.

Our eventual goal is to apply the method to the output from a network
of detectors, which will provide an even better ability to
discriminate signal and noise, reconstruct the signal, plus allow us
to estimate the source position on the sky. This would necessitate the
use of catalogs for stellar core collapse waveforms which are both
quantitatively more extensive and qualitatively improved, so that the
parameter space of likely events can be completely covered and very
finely resolved. For instance, a larger selection of available
equation of state tables will extend the phase space in that
respective direction, while a more accurate inclusion of neutrino
effects will both improve the quality of the burst signal from core
bounce and as well yield a much more appropriate characterization of
the low-frequency signal from post-bounce convective bulk motion in
the post-shock region and in the proto-neutron star. In addition,
other prospective emission mechanisms for gravitational waves in a
core collapse event, like proto-neutron star pulsations or
nonaxisymmetric rotational instabilities, could be also considered
(for a comprehensive review on various such mechanisms,
see~\cite{Ott2009}).

The method presented in this paper will provide a better way to produce 
statistical and probabilistic statements about the physical parameters 
associated with a stellar core collapse event, and the ability to make 
statements about actual signals observed by interferometric detectors.
The data from multiple interferometers will be coherently
analyzed; this will allow the estimation of parameters associated with
the location of the source on the sky, similar to what can be achieved
with coherent MCMC analysis of binary inspiral gravitational wave
data~\cite{RoeverMeyerChristensen2007a, RoeverEtAl2007b}. As noted
previously in our examples, the timing accuracy is better than
$ 1 \mathrm{\ ms}$ for a single interferometer observation; 
a multiple interferometer coherent MCMC should provide 
very good sky localization estimation,
and the use of principal components
within the framework of Searle et al.~\cite{SearleSuttonTinto2009}
also shows great promise.
In addition, a coherent analysis will also allow us to
better distinguish mismatch ($ \bm $, common to all detectors) from
noise ($ \bepsilon $, different between different detectors) and hence
should improve the waveform reconstruction. We also plan to
incorporate a flexible model for the detector noise spectrum, as
described in~\cite{RoeverMeyerChristensen2008}. A proper accounting
for the noise spectral densities of the interferometers could also be
applied in producing the PCA eigenvectors and eigenvalues, i.e.\ these
would not simply be the result of a least-squares fit, but rather a
noise-weighted least-squares fit. The resulting basis vectors should
then better reflect what is actually ``visible'' to an interferometric
detector within its limited sensitivity band.

An important part of our long-term research program will be to ensure 
that the catalog of waveforms truly spans and encompasses those from 
physically possible rotating stellar core collapses; it will be 
critical for us to consult closely with experts in the field 
\cite{Ott2009,OttEtAl2007,DimmelmeierEtAl2007, DimmelmeierEtAl2008a} 
to assure that the physical parameters for the waveforms we will use 
cover the range of natural possibilities. Typically when initially 
constructing the initial models one  already has some idea about how 
many intermediate steps per free parameters are required. When the 
waveforms are generated as the output from the calculations (along 
with the other output data), then the variations of the waveforms and 
the other data give some indication if the original choice of model 
construction and parameter division are sensible. While the signal 
catalog needs to stretch over the extreme limits of the values for 
the physical parameters, a strength of our method is that the table 
need not be densely populated (as is the case with the template bank 
for coalescing binary signals); this the advantage provided by the use 
of the PCA\@. For example, when the PCA eigenvalues are derived, their 
variation provides guidance about the quality of our catalog. We can 
test whether the waveform table contains enough entries by successfully 
reconstructing additional (off-table) signals placed within the 
parameter space; this was essentially the method used for verifying the 
results presented in this article.

The technique described in this paper could prove to be advantageous
for signal reconstruction and parameter estimation when numerical
techniques are required in order to produce the waveforms. For example,
we can imagine our method as being useful for estimating parameters
for complicated binary inspiral signals; numerical relativity
calculations are producing a wide array of complicated inspiral
signals~\cite{AylottEtAl2009a}. 
We have already shown
in Section~\ref{section:correspondence} that the fitted PC
coefficients can be related to the original physical parameters of the
signals in the catalog. Future research will explore how the joint
posterior distribution of the PC coefficients can be employed for
statistical inference on the physical parameters by linking the different 
parameter spaces via Procrustes rotation~\cite{MardiaKentBibby}.

This method should also prove to be useful in distinguishing a real
stellar core collapse event from a common noise \textsl{glitch} in
the data. As displayed in Fig.~\ref{fig:CatalogCoefs1}, there is a
characteristic pattern to be found with PCA values associated with a
\textsl{real} signal. A glitch produced by noise, when reconstructed
via PCA, will likely fall outside the pattern formed by stellar core
collapse signals. Alternatively, one could set up an alternative
glitch model, which, instead of using numerical simulations for the
(PCA) basis vector generation, is trained on sets of measured
waveforms that are known to be instrumental artifacts. We intend to
test this potential \textsl{veto} technique.

In this paper we have reduced the table of 128 waveforms to 10 or 20
of the most important PC eigenvectors. Instead of arbitrarily choosing
the number of eigenvectors to use, we plan to have our MCMC optimize
the signal reconstruction through the use of a reversible jump MCMC,
similar to the method described in~\cite{UmstaetterEtAl2005a}, thereby
treating the number or subset of PCs as another unknown. In this
Bayesian approach the optimal number of eigenvectors is determined by
weighing the benefit of a good signal fit with a large number of
eigenvectors against the Ockham's Razor penalization when the model
gets overly complex.


\acknowledgments

It is a pleasure to thank Christian D.\ Ott and Patrick J.\ Sutton
for a careful reading of the script. H.D.\ acknowledges a Marie Curie
Intra-European Fellowship within the 6th European Community Framework
Programme (IEF 040464). This work was supported by 
the Max-Planck-Society,
The Royal Society of New Zealand Marsden Fund grant \mbox{UOA-204}, 
National Science Foundation grants \mbox{PHY-0553422} and \mbox{PHY-0854790},
the Fulbright Scholar Program,
the DAAD and IKY (IKYDA German--Greek research travel grant), 
by the European Network of Theoretical Astroparticle Physics ENTApP
\mbox{ILIAS/N6} under contract number \mbox{RII3-CT-2004-506222},
and by the Science and Technology Facilities Council 
of the United Kingdom and the Scottish Universities Physics Alliance.


\appendix

\section{Definition of discrete Fourier transform}
\label{appendix:dft}

The discrete Fourier transform is defined for a real-valued function
$ h $ of time $ t $, sampled at $ N $ discrete time points at a
sampling rate of $ \frac{1}{\Delta_t} $.

The transform maps from
\begin{equation}
  \label{eqn:DFTtimedomain}
  \{ h (t) \in \realline:
  t = 0, \Delta_t, 2 \Delta_t, \ldots, (N - 1) \Delta_t \}
\end{equation}
to a function of frequency $ f $
\begin{equation}
  \{ \tilde{h} (f) \in \complexnumb:
  f = 0, \Delta_f, 2 \Delta_f, \ldots, (N - 1) \Delta_f \},
\end{equation}
where $ \Delta_f = \frac{1}{N \Delta_t} $ and
\begin{equation}
  \label{eqn:DftDefinition}
  \tilde{h} (f) = \sum_{j = 0}^{N - 1} h (j \Delta_t)
  \exp (- 2 \pi \imag j \Delta_t f).
\end{equation}
Since $ h $ is real-valued, the elements of $ \tilde{h} $ are
symmetric in the sense that $ \tilde{h} (i \Delta_f) $ and
$ \tilde{h} ((N - i) \Delta_f) $ are complex conjugates; this allows
to restrict attention to the non-redundant elements indexed by
$ i = 0, \ldots, N / 2 $.


\section{Fourier domain model}
\label{appendix:fourier_domain}

A Gaussian distribution of a (time-domain) random vector also implies
gaussianity for the Fourier transformed vector. In particular, if
$ n (t) $ is (zero mean, stationary) Gaussian noise with one-sided
power spectral density $ S_1 (f) $, then in the limit of large $ N $
and small $ \Delta_t $ the real and imaginary components of the
discrete Fourier transformed time series are independently zero-mean
Gaussian distributed:
\begin{equation}
  \label{eqn:FDomainGaussian}
  \prob\bigl(\re ( \tilde{n} (f_j) ) \bigr) = \normaldistn (0, \sigma_{f_j}^2),
  \quad
  \prob\bigl( \im (  \tilde{n} (f_j) ) \bigr) = \normaldistn (0, \sigma_{f_j}^2),
\end{equation}
where the variance parameter is
$ \sigma_{f_j}^2 = \frac{N}{4 \Delta_t} S_1(f_j) $~\cite{Finn1992,
  Brillinger, RoeverMeyerChristensen2008}.


\section{Joint and conditional posteriors}
\label{appendix:posteriors}

For the random effects model from Sec.~\ref{subsection:model_2}
the joint distribution of data and parameters may be factored out
into:

\begin{equationarray}
  & & \pdf (\by, \bbeta, T, \sigma_m^2, \tilde{\bm})
  \nonumber
  \\
  & = & \pdf (\by | \bbeta, T, \sigma_m^2, \tilde{\bm})
  \times \pdf (\tilde{\bm} | \bbeta, T, \sigma_m^2)
  \nonumber
  \\
  & & \times \pdf (\sigma_m^2 | \bbeta, T) \times \pdf (\bbeta, T)
  \\
  & \propto & \exp \biggl( - 2 \sum_{j = 1}^{N / 2}
  \frac{\frac{\Delta_t}{N} | \tilde{y}_j -
  \bigl( \tilde{\bX}_{\!(T)} \bbeta \bigr)_j -
  \tilde{m}_j |^2}{S_1(f_j)} \biggr)
  \label{eqn:jointLikeli1}
  \\
  & & \times \bigl( \sigma_m^2 \bigr)^{-N / 2}
  \exp \Bigl( - \sum_{i = 1}^{N} \frac{m_i^2}{2\sigma_m^2} \Bigr)
  \label{eqn:jointLikeli2}
  \\
  & & \times \bigl( {\textstyle\frac{1}{N} \| \bX\bbeta \|^2} \bigr)^{\frac{\nu}{2}}
  \bigl( \sigma_m^2 \bigr)^{- 1 - \frac{\nu}{2}}
  \exp \Bigl( \frac{ - \nu \, \gamma \, {\textstyle\frac{1}{N} \| \bX\bbeta \|^2}}{2 \sigma_m^2} \Bigr), \qquad
  \label{eqn:jointLikeli3}
\end{equationarray}%
where the proportionality here refers to keeping the data and known
parameters constant. In the above equation, the first
term~(\ref{eqn:jointLikeli1}) is the `usual' time-domain likelihood
from Appendix~\ref{appendix:fourier_domain}, the
term~(\ref{eqn:jointLikeli2}) is the likelihood of the mismatch
(random effect), and the term~(\ref{eqn:jointLikeli3}) is a conditional
prior of the mismatch variance parameter. By fixing the values for
different subsets of parameters, one can determine
\textsl{conditional} posterior distributions that are useful for the
Gibbs sampling implementation (see Section~\ref{subsection:mcmc}).


\bibliography{/home/christian/literature/literature}

\end{document}